\documentstyle[emulateapj]{article}

\submitted{To appear in the Astrophysical Journal}

\input epsf.sty

\newcommand{\etal}{et~al.\ }
\newcommand{\eg}{e.g.\ }
\newcommand{\ie}{i.e.\ }
\newcommand{\Msun}{M$_{\odot}$}

\newcommand{\kms}{km~s$^{-1}$}
\newcommand{\Mej}{M$_{\rm ej}$}
\newcommand{\HeI}{He~{\sc i}}

\newcommand{\OI}{O~{\sc i}}

\newcommand{\NaI}{Na~{\sc i}}

\newcommand{\SiII}{Si~{\sc ii}}

\newcommand{\CaII}{Ca~{\sc ii}}

\newcommand{\FeII}{Fe~{\sc ii}}

\newcommand{\CoII}{Co~{\sc ii}}
\newcommand{\NiII}{Ni~{\sc ii}}

\newcommand{\Cofs}{$^{56}$Co}
\newcommand{\Nifs}{$^{56}$Ni}

\begin{document}

\title{A Spectroscopic Analysis of the Energetic Type Ic `Hypernova'
SN~1997ef}

\author{Paolo A. Mazzali\altaffilmark{1,2}, Koichi Iwamoto\altaffilmark{3}, 
Ken'ichi Nomoto\altaffilmark{1,4} }

\altaffiltext{1}{Research Centre for the Early Universe, School of
Science, University of Tokyo, Bunkyo-ku, Tokyo 113-0033, Japan}
\altaffiltext{2}{Osservatorio Astronomico, Via Tiepolo, 11, I-34131 Trieste, 
Italy}
\altaffiltext{3}{Department of Physics, Nihon University, Chiyoda-ku,
Tokyo 101-8308, Japan}
\altaffiltext{4}{Department of Astronomy, School of Science,
University of Tokyo, Bunkyo-ku, Tokyo 113-0033, Japan }

\begin{abstract}

The properties of the bright and energetic Type~Ic SN~1997ef are investigated
using a Monte Carlo spectrum synthesis code. Analysis of the earliest spectra
is used to determine the time of outburst. The changing features of the
spectrum and the light curve are used to probe the ejecta and to determine
their composition, verifying the results of explosion calculations. Since
synthetic spectra computed using our best explosion model CO100 are only
moderately good reproductions of the observations, the inverse approach is
adopted, and a density structure is derived by demanding that it gives the best
possible fit to the observed spectrum at every epoch analysed. It is found that
the density structure of model CO100 is adequate at intermediate velocities
(5000--25000 \kms), but that a slower density decline ($\rho \propto r^{-4}$)
is required to obtain the extensive line blending at high velocities
(25000--50000 \kms) which is the characterising feature of this and other
energetic type Ic Supernovae. Also, the inner `hole' in the density predicted
by the model is found not to be compatible with the observed evolution of the
spectrum, which reaches very low photospheric velocities at epochs of about 2
months. The `best fit' density distribution results in somewhat different
parameters for the SN, namely an ejecta mass of 9.6\Msun\ (v. 7.6\Msun\ in
CO100) and an explosion kinetic energy of $1.75 \cdot 10^{52}$erg (v. $8 \cdot 
10^{51}$erg in CO100). This revised value of the kinetic energy brings
SN~1997ef closer to the value for the `prototypical' type Ic `hypernova'
SN~1998bw. The abundance distribution of model CO100 is found to hold well. 
The modified density structure is used to compute a synthetic light curve,
which is found to agree very well with the observed bolometric light curve
around maximum. The amount of radioactive $^{56}$Ni produced by the SN is
confirmed at 0.13\Msun. In the context of an axisymmetric explosion, a somewhat
smaller kinetic energy than that of SN~1998bw may have resulted from the non
alignment of the symmetry axis of the SN and the line of sight. This might also 
explain the lack of evidence for a Gamma Ray Burst correlated with SN~1997ef. 
\end{abstract} 


\section{Introduction} 

SN~1997ef in UGC4107 was recognised as a peculiar and interesting object as
soon as its first spectra were taken by the Harvard-CfA team (Garnavich \etal
1997a,b,c). The spectra displayed very broad features, quite unlike those of
any other SN known to date, so that it was not even clear whether what was
observed was an absorption or an emission spectrum. Continued observation
revealed a light curve typical of a SN deriving from a compact object (a SN~Ia
or a SN~Ic), but much broader than the templates for both of these classes.
Later spectra showed more resolved spectral lines, clarifying that the broad
features were actually very extended absorption blends, and that the SN was
therefore in the photospheric epoch. Lines of \CaII, \OI\ and \FeII\ were 
strong, and the \SiII\ 6347, 6371\AA\ line appeared to be comparatively weak,
so the SN was classified as type Ic, as also supported by the overall
similarity with the spectra of other SNe~Ic such as SN~1994I (Filippenko 1997; 
Millard \etal 1999).

Typical models for a low mass, low kinetic energy SN~Ic (Nomoto \etal 1994; 
Iwamoto \etal 1994) gave good fits to the light curve, but the synthetic
spectra computed using such models, while yielding the correct types of
spectral lines, completely failed to reproduce the observed large line width,
which we estimated as at least 20000\kms for some of the strongest lines
(Iwamoto \etal 2000, hereafter Paper I).

Soon after SN~1997ef, however, the extraordinarily bright and powerful type Ic
SN~1998bw was discovered as the optical counterpart of GRB980425 (Galama \etal
1998). A great deal of study, both observational and theoretical, was and still
is being devoted to this SN, which was soon realised to be a massive type Ic
event of exceptionally large kinetic energy ($KE \sim 3 \cdot 10^{52}$ ergs,
Iwamoto \etal 1998; Woosley, Eastman \& Schmidt 1999; Branch 2000), where the
link with the GRB may lie in a rather asymmetric explosion (MacFadyen \&
Woosley 1999, H\"{o}flich \etal 1999, Khokhlov \etal  1999, Maeda \etal 2000).
It was only natural then also to analyse SN~1997ef as a powerful `hypernova'.
Paper I presents a detailed study of the light curve and the spectrum using two
models, one for a `normal' SN~Ic (model CO60: \Mej = 4.4\Msun, KE = 10$^{51}$
erg) and another for a `hypernova' (model CO100: \Mej = 7.6\Msun, $KE = 8 \cdot
10^{51}$ erg). Both models have a \Nifs\ mass of 0.15\Msun. In Paper I we
showed that both models reproduce the light curve reasonably well, but only the
more massive and energetic model CO100 can also produce reasonable synthetic
spectra, and therefore they opted for the `hypernova' model and suggested that
SN~1997ef may be a somewhat less extreme case than SN~1998bw. A possible
connection with a GRB has been searched for in archives for SN~1997ef. The
result was that GRB971115 could be compatible with SN~1997ef in position and
time of occurrence (Wang \& Wheeler 1998), but the statistical significance of
the correlation is much weaker than for SN~1998bw and GRB980425.

In this paper, we concentrate on the spectroscopy of SN~1997ef. This is
interesting in many respcts. First, it is important that we not only fit the
spectra and their evolution using one explosion model, but that we also
understand which lines are present in the spectrum. Since we have available a
sufficiently large number of observed spectra, we can follow the spectral
evolution and thus probe different depths in the ejecta and verify the
composition as a function of velocity, and also describe how the spectrum
changes as a function of luminosity as the SN moves along the light curve. We
use for this purpose the Monte Carlo (MC) code originally described in Mazzali
\& Lucy (1993) and further improved and discussed by Lucy (1999) and Mazzali
(2000). The code includes the effect of photon branching, which must be very
important in a situation where the high expansion velocities lead to extensive
line blanketing. Also, since the code uses the bolometric luminosity as input,
we can compare the required $L$ values with those of a light curve calculation
relaxing the assumption $BC=0.0$ made in Paper I and thus improve the
comparison between model $L_{\rm Bol}$ from light curve calculations and
observed $V$ mags.

Furthermore, we use spectrum synthesis to date the earliest spectra, and thus
to determine with some accuracy the time of explosion, which is important for
both the GRB connection and the light curve calculation. 

Finally, having noticed that even synthetic spectra obtained from CO100 do not
give wonderfully good fits to the observed spectra, we adopt the `inverse'
approach and try to determine the density and abundance distribution with
velocity by looking for a best fit to the spectra. This method was pioneered by
Branch (2000), who suggested for SN~1997ef a much flatter density dependence
($\rho \propto r^{-2}$) than what is predicted by model CO100 (where the
density index is $n=-7$. Consequently, for SN~1997ef Branch (2000) obtains a
much larger ejecta mass (6\Msun\ above $v = 7000$ \kms) and kinetic energy 
($3 \cdot 10^{52}$erg above the same velocity). 

Since we study several spectra we can control $\rho$ as a function of $v (=
r/t)$, and hence build our own $\rho(r)$. We do confirm that the spectra
require a flatter density dependence, but we show that this is only necessary
at high velocity ($v > 25000$ \kms). Examining spectra at advanced epochs, we
also show that the predicted density `hole' at low velocities is not compatible
with observations. We therefore derive new values for both \Mej and $KE$. 

We take this process one step further by computing a synthetic light curve from
our derived $\rho(r)$, and show that this gives a very good fit to the observed
light curve of SN~1997ef. 

In the rest of this paper we first describe how we proceeded to date the
spectra, then discuss the analysis of 6 spectra, reviewing the results and
their implications. Then we review the properties of the model for SN~1997ef as
we derived it, and discuss how it differs from model CO100, which was computed
to fit the light curve, not the spectra. We also show the synthetic light curve
obtained from our ad hoc model for $\rho(r)$. Finally, we review and discuss
our findings, including possible reasons why SN~1997ef was a hypernova but no
GRB was apparently linked to it.

\section{Dating the spectra} 

We have selected for modelling 6 spectra of SN~1997ef obtained by the
Harvard-CfA group (Garnavich \etal in prep.; see Paper I, Figs. 8 and 9). We
have chosen well exposed, high S/N spectra, with a large wavelength coverage,
and tried to sample the light curve as uniformly as possible. The first three
spectra are from epochs around maximum (which was not observed either
spectroscopically or photometrically but was probably reached around Dec 10,
1997 at $V \sim 16.5$. The dates of the spectra are Nov 29, which is very soon
after discovery, Dec 5 and Dec 17. All the spectra show very broad lines. They
are all early enough that they are sensitive to changes in the kinetic energy
and to the assumed epoch of outburst. The next two spectra have dates Dec 24
and Jan 1, which is around the time when the SN enters the tail of the light
curve. The lines in these spectra becoming progressively narrower, indicating
that the high velocity part of the ejecta is becoming optically thin. The final
spectrum has date Jan 26, 1998, which is well on the light curve tail. Line
velocities are quite small, suggesting that the spectrum is formed deep in the
ejecta. One of the challenges for a good light curve calculation is to follow
the evolution of the photospheric velocity with time. Even model CO100 was not
perfect in this respect (cf. Paper I, Figure 7).

Since the epoch of maximum was not observed, light curve models are not very 
tightly constrained. Since the rise time to maximum depends on the structure of
the model and the abundance distribution, the range of allowed parameters could
be effectively costrained if at least the time of explosion could be
determined, albeit with some uncertainty. This could be attempted using
spectrum synthesis and the measured photospheric velocity as determined in
Paper I from the velocity of the \SiII\ line. It could be expected that this
line is formed somewhat above the photosphere at very early epochs, like in
SNe~Ia (\eg  SN~1994D, Patat \etal 1996).

Since our synthetic spectra are computed using the luminosity $L$, the epoch
$t$ and the photospheric velocity $v_{\rm ph}$ as input, an appropriate value
of $t$ must be found that gives a photospheric radius $R_{\rm ph} = v_{\rm ph}
t$ which, when combined with $L$, produces temperatures in agreement with the
observed spectrum. In particular, although the continuum is not well determined
in the very complex spectra of SN~1997ef, spectral lines serve a double role:
they depend on the temperature (and hence on $R_{\rm ph}$ and $L$) through
their ratios, and on $v_{\rm ph}$ through their displacement. Not many lines
are visible in the earliest spectra, but at least \SiII\ 6346, 6371\AA\ and the
broad complex which can most likely be identified as \OI\ 7774\AA, \OI\
8447\AA\ and the \CaII\ IR triplet could be used to guide the calculation of
the synthetic spectra. 

Models for Nov 29 show that acceptable fits are found for $t = 7-9$ days and
$v_{\rm ph} = 18000-15500$ \kms. Greater epochs require smaller velocities, so
that $R_{\rm ph}$ always has a value of about $1.1 \cdot 10^{11}$cm. This
radius, combined with the value of $L$, which depends almost entirely on the
assumed distance, gives a temperature appropriate to get a good overall fit to
the spectrum. If we assume \eg $t = 11$ days, then we find a good overall fit
to the spectrum for $v_{\rm ph} = 13500$ \kms, which is too small and is
reflected on the line shift. The opposite happens if $t \leq 7$ days. In order
to strenghten our result, we applied the same technique to the Dec 5 spectrum,
and we found acceptable fits for $t = 14-15$ days and $v_{\rm ph} =
10000-9500$\kms. Therefore, the two epochs give consistent results, and we can
estimate the time of outburst to have been Nov 20--22. Since the fits obtained
for $t = 9$ days on Nov 29 and $t = 15$ days on Dec 5 were particularly good,
we selected Nov 20 as our reference epoch, also for the comparison with the
light curve and the  photospheric velocities. We will show the fits in the next
sections, when discussing the various epochs. In Figure 1 we show as an example
the Nov 29 spectrum and two synthetic spectra computed as above for $t = 9$
days (dashed line) and $t = 11$ days (dotted line), respectively, using model
CO100. The difference in velocity is clearly visible.

These results depend on the selection made for two other quantities, distance
and overall ejecta mass, as discussed by Mazzali \& Schmidt (2000) in the
context of Type Ia SNe.  For all models, we adopted a distance modulus of
33.63, \ie a distance of 53.1 Mpc, as estimated from a recession velocity of
3450\kms (P. Garnavich, 2000, private communication) and $H_0 = 65$\kms
Mpc$^{-1}$. We assume no extinction ($E(B-V) = 0.0$, which is justified by the
absence of a narrow interstellar \NaI D line from the spectrum of SN~1997ef at
all phases. As for the ejected mass, we already showed in Paper I that this
must be large in order to fit both the light curve and the spectra. Therefore
in this paper we use the solution found in Paper I, model CO100, as a
reference.

\section{The 29 November 1997 spectrum} 

This is to our knowledge the earliest spectrum available of SN~1997ef. Although
it is somewhat noisy and it does not extend very far to the red, it
nevertheless includes most of the main features, which made SN~1997ef such a
peculiar and interesting object. The spectrum (Figure  2) has essentially no
continuum, and it is characterised by four broad minima, at roughly 3900, 4700,
6000 and 7500\AA. The minima are separated by three rather sharp peaks, at
4300, 5200 and 6300\AA, respectively. The minima are so broad and the peaks so
strong that the spectrum could be confused for an emission spectrum. Both
Branch (2000) and Paper I showed that the real nature of the spectrum is a
superposition of absorption lines. We already pointed out that the `emission
peaks' are the result of photon travel in a medium of large line opacity.
Photons redshift their way through the envelope until they find a portion of
the spectrum where line opacity is low because of the intrinsic distribution in
wavelength of the atomic transitions. We call such a region a `line-free
window'.  A large number of photons escape through these regions - all the
photons coming from the optically thick region immediately to the blue, and
this leads to a large flux and to the observed `peaks'. While the width of the
peak depends solely on the width of the line-free window, its strength depends
both on the flux at the lower boundary (the photosphere) and on the width of
the blanketing region which feeds photons to it: the broader the blanketing
region, the more photons must reach the window to escape, resulting in a higher
peak flux. This can be easily seen by inspection of the spectrum.

In Paper I we showed that a synthetic spectrum based on model CO100 gives a
reasonably good reproduction of the observations, and allows us to identify the
four absorptions as follows: 3900\AA: \CaII\ H\&K + \NiII, \CoII\ and \SiII\
lines; 4700\AA: mostly \FeII\ lines; 6000\AA: \SiII\ 6347, 6371\AA\ plus weaker
\SiII\ lines and \NaI D further to the blue; 7500\AA: \OI\ 7774\AA. In spectra
computed with model CO100 the \OI\ 7774\AA\ and the \CaII\ IR triplet are well
separated, but unfortunately the observed spectrum does not extend far enough
to the red to show the \CaII\ region. With reference to that synthetic
spectrum, which is shown again here as the dashed line in Figure 2, we notice
that none of the absorptions extend quite far enough to the blue.  For the two
bluer troughs, which are blends of very many lines, this may possibly be due to
the lack of lines resulting from neglecting some element or not using a
sufficiently complete line list, but this seems unlikely for the two redder
troughs, where a few very strong lines dominate.

The synthetic spectrum has $t = 9$ days, $\log L = 42.13$ (erg s$^{-1}$),
$v_{\rm ph} = 15500$ \kms. The observed $V$ magnitude (16.7) and the synthetic
one (16.75) agree quite well, but the spectrum has a large Bolometric
Correction (0.28), which indicates that the assumption $BC = 0.0$ made in Paper
I when fitting the light curve is at least weak. The observed \SiII\ line
velocity is 19000 \kms, and the core of the \SiII\ absorption is well
reproduced even using model CO100. Therefore, in order to improve the synthetic
spectrum, we modified the density only above 20000 \kms. While the original
CO100 model has $\rho \propto r^{-8}$ in the outer part, we made the outer
density radial dependence flatter. As a consequence, high velocity absorption
becomes stronger. Of course, this happens first in the strongest lines, the
\CaII\ H\&K doublet and IR triplet first and foremost. The strength of these
lines increases dramatically, because they come from low excitation levels
which are still highly populated even at the very low densities and
temperatures of the outer envelope ($T_e \sim 4000$K), and so these lines
increase greatly in width. Next come the \SiII\ and \OI\ lines. For these
lines, the drop in the level population with radius is much steeper, since they
come from levels on average about 10eV above the ground state, and so they do
become broader, but not so noticeably, and their cores do not shift much
towards the blue. All other lines, strong and weak are similarly affected, to a
higher or lesser degree depending on the ionisation and excitation of the
levels.

Since we wanted to introduce as little change from model CO100 as possible, we
choose the steepest density profile which gave a good fit to the spectrum. This
was obtained with a density power law index $n= -4$ at $v > 25000$\kms. When
this change is introduced the \CaII\ H\&K doublet, the \SiII\ line and the \OI\
7774\AA\ line broaden to the point that the observed absorption troughs are
very well reproduced. Further to the red, in the unobserved part of the
spectrum, the \CaII\ IR triplet is predicted to broaden and blend with \OI\ 
7774\AA. The contribution of \OI\ 8446\AA, which is also a strong line, is
nevertheless small compared to that of \CaII. The blending of \OI\ 7774\AA\ and
the \CaII\ IR  triplet is an important discriminant for the density structure
at high velocities. Fortunately, the Dec. 5 spectrum covers that region. The
blue sides of the peaks at 4300 and 5000\AA\ are also fitted much better, as
the increased density causes the emission to set in at higher velocities.

The only region where no improvement is seen is near 4500\AA. The \FeII\ and
\CoII\ lines causing the absorption near 4750\AA\ are not sufficiently strong
to extend to the blue, and there are not enough strong lines that absorb near
4400--4500\AA. It is possible that we are either missing lines or ignoring
some relevant element. Interestingly, Branch (2000), who used a code that is
very efficient for identifying lines, seemed to have the same problem.

The modification to the outer density profile has only a small influence on
the total mass, since it adds only about 0.25\Msun\ in the outer ejecta, but
it does lead to a significant increase in the overall kinetic energy, since all
the mass is added to the high velocity shells, where it represents a dramatic
increase, of 1--2 orders of magnitude. The kinetic energy above 20000 \kms\ 
increases from $10^{51}$ to about $10^{52}$erg, and the total kinetic energy 
increases from $8 \cdot 10^{51}$ to $1.75 \cdot 10^{52}$erg. This is quite a
large change, but it is still smaller than the value adopted for SN~1998bw.
Also, both the outer slope and the revised $KE$ are, respectively, steeper and
smaller than the values suggested by Branch (2000). We will discuss the
implications of these revised values later. An increase in $KE$ only makes
SN~1997ef more like a hypernova.

The best fitting ad hoc model has the same input parameters as the spectrum 
computed using CO100. The increased outer density makes the temperature only
slightly higher. The synthetic spectrum has $B=17.45$, $V=16.74$ and $BC =
0.29$, so that $M_{Bol} = -16.70$. Therefore comparing $M_{Bol}$ with $V$ is
misleading, since $V$ rises more rapidly than $M_{Bol}$.

Finally, a word of comment about the presence of He in the spectrum. In
SN~1998bw, \HeI\ 10830\AA\ was almost certainly observed in absorption at early
epochs (Patat \etal 2000). Its presence could be understood if the He shell was
not completely lost before the explosion. Unfortunately, IR spectra are not
available for SN~1997ef, so the best place to look for He is \HeI\ 5868\AA.
This line is very close to \NaI D, which is rather strong and contributes to the
broad 6000\AA\ trough with an absorption near 5450\AA. A small He mass does not
produce a strong line unless very large departure coefficients are assumed, and
in any case a strong \HeI\ line is not necessary to fit the spectrum.
Therefore, although we cannot be final, we do not think \HeI\ 5868\AA\ is
present in the Nov 29 spectrum.

\section{The 5 December 1997 spectrum}

The second spectrum in our analysis has a good S/N and a very good wavelength
coverage, extending from 3100 to 8700\AA. It is actually quite similar to the
29 Nov. one, but more detail is visible (Figure 3), \eg the structure of the
complex 6000\AA\ absorption and, very importantly, the \CaII\ IR triplet is
visible, and clearly blended with \OI\ 7774\AA, which is a strong argument in
favour of a `flat' outer density distribution, as discussed above. That the
feature measured at 7500\AA\ is a blend is clear, given the presence of two
distinct absorptions, whose wavelengths match those of the two component
multiplets, as shown also by the spectrum computed with model CO100. 

The synthetic spectrum based on model CO100 was shown in Paper I and again here
as the dashed line in Figure 3. It has parameters $t = 15$ days, $\log L =
42.15$ (erg s$^{-1}$), $v_{\rm ph} = 9500$\kms. Both $L$ and $R_{\rm ph}$ have
values similar to those on Nov. 29, therefore the temperatures are similar and
so are the two spectra, although the photosphere is quite a bit deeper on Dec.
5. This is shown also by the \SiII\ line velocity, which is reduced to
13000\kms. The observed $V$ mag on Dec 5 was 16.5, and the CO100 spectrum
reproduces it well ($V = 16.63$). The problem with that model is essentially
the same as that discussed for the Nov 29 model: all troughs do not extend far
enough to the blue. Furthermore, \OI\ 7774\AA\ and the \CaII\ IR triplet do not
blend anywhere near as much as they should when the original model is used. In
that spectrum the \CaII\ IR triplet extends only to about 7850\AA, \ie a \CaII\
velocity of $\sim 25000$\kms. If the \CaII\ IR triplet is to blend with the
\OI\ line, it must reach at least 7650\AA\ in the blue wing, \ie $v \sim 
31000$\kms. This is consistent with the modified density structure adopted for
the Nov 29 spectrum. Since $v_{\rm ph}$ is much lower than the values at which
the density has been modified ($v > 20000$\kms), we can use the same density
structure as for Nov 29.

The spectrum we computed for the modified structure and the same input
parameters as above is shown in Figure 3. The improvements are apparent, and
they concern mostly the \CaII\ IR triplet + \OI\ 7774\AA\ feature in the red
and the blue region near 4000\AA, which is impoved by the strengthening and
broadening of \CaII\ H\&K. Improvements in the trough at 4700\AA\ are again
only marginal, as are those to the \SiII\ line region. Surprisingly perhaps,
the \SiII\ line does not become broader, but this can be understood since the
high velocity regions where the density has been enhanced are far above the
photosphere and are of too low density now to cause much \SiII\ absorption. The
situation is made more complicated by the apparent development of an absorption
shoulder near 5900\AA. We cannot offer a clear identification for this
shoulder, except for several lines of \OI\ at 6157\AA, but these are not very
strong. The shoulder was not present on Nov 29, nor is it on Dec 17. Another,
stronger shoulder is visible near 5750\AA. This is where \SiII\ 5958, 5979\AA\
falls, and the synthetic spectra show that this doublet, combined with \NaI D
further to the red, is strong enough to produce a shoulder stronger than the
observed one without resorting to \HeI. Therefore we can confirm the
non-detection of \HeI\ 5876\AA. Note that the Si abundance in this model was
lower than on Nov 29. This was necessary to fit the core of the \SiII\ 6347,
6371\AA\ line. Even the modified density does not give a perfect synthetic
spectrum, but the excellent behaviour in the red confirms that we are adopting
the correct measures when using an extended outer envelope.

If a density law flatter than $n = -4$ is adopted in an attempt to improve the
\SiII\ line two problems emerge: 1) \OI\ 7774\AA\ extends too far to the blue
and 2) the increasing strength of \CaII\ H\&K causes too much blocking of the
UV spectrum and, consequently, a synthetic peak at 4500\AA\ much stronger than
what is observed. These problems would be even worse in the Nov 29 spectrum,
and so we chose to retain the $n = -4$ solution as a compromise and suggest
that some alternative line may be the reason for at least the shoulder at
5900\AA. 

Our best-fitting spectrum has $B = 17.53$, $V = 16.58$ and again a large $BC =
0.40$, so that $M_{Bol} = -16.75$. The increased $BC$ is apparent if one looks
at the ratio of the two main peaks, at 4300 and 5200\AA. Therefore, $M_{Bol}$ 
is significantly smaller than $V$ at this epoch, which is near maximum in the
$V$ light curve but appears to be still very much on the rising branch of the
bolometric curve.

\section{The 17 December 1997 spectrum}

This third spectrum has again a limited wavelength coverage, but it is the
first one available after Dec 5, and it is therefore useful to study the
evolution of the SN immediately following $V$ maximum. The spectrum, which is 
displayed in Figure 4 has $V = 16.6$. The overall aspect of the spectrum is
similar to the previous two epochs, which indicates that the temperature
conditions have probably not changed greatly. 

However, closer inspection, and in particular a wavelength comparison of this
and the two earlier spectra, reveals that all the main absorption and emission
peaks have shifted to the red by 100-200\AA, depending on the wavelength. A
similar, but smaller shift also occurred between Nov 29 and Dec 5. The
immediate interpretation is that we are observing the effect of the inward
motion of the photosphere through the ejecta as they expand and thin out. Note
that this recession takes place essentially in the lagrangian mass/velocity
frame, and not in the observer's (radius) frame, much like in the Plateau phase
of some SNe~II. The \SiII\ velocity measured from this spectrum is in fact only
8000\kms. The change in line absorption velocity is larger between Dec 5 and
Dec 17 than it was between Nov 29 and Dec 5 because the time interval between
the spectra is larger. All absorptions move by about the same amount, and the
emissions move as well. This is obvious given our explanation for what the
emissions really are: if the blanketed regions move, the opacity windows must
move along with them, just like in P-Cygni profiles caused by a single line, a
situation of which the case of SN~1997ef represents after all just an extreme
extension. Looking in more detail, the shoulder which affected the \SiII\ line
near 5900\AA\ on Dec 5 seems to have disappeared on Dec 17, leaving serious
questions about its identity. The absorption near 5000\AA\ now shows more
structure, with minima at 4800 and 4950\AA\ and two shoulders at 5100 and
5300\AA, which resembles quite closely the typical profile observed in SNe~Ia
near maximum and soon thereafter. In SNe~Ia the entire feature is attributed
mostly to \FeII\ lines, and this was confirmed in SN~1997ef on both Nov 29 and
Dec 5.

The synthetic spectrum computed with CO100 is shown as the dashed line in
Figure 4. The structure in the \FeII\ absorption is reproduced reasonably well,
and the width problem now does not affect either the feature at 4300\AA\ or
that near 7500\AA. This is probably because the outer high velocity regions are
so far above the photosphere now that they do not really cause any significant
absorption, and so the original model CO100 gives an acceptable description of
the density structure in the intermediate velocity regions. Unfortunately this
spectrum does not cover the \CaII\ IR triplet, which is predicted not to blend
with \OI\ 7774\AA\ in the CO100 model. The synthetic spectrum was computed with
the following parameters: $t = 27$ days, $\log L = 42.20$ (erg s$^{-1}$), 
$v_{\rm ph} = 7500$\kms. The temperatures are lower than on Dec 5 but
comparable to those on Nov 29, and $R_{\rm ph}$ is actually larger than on both
previous epochs, although the mass above the photosphere is now as large as
5\Msun. Compared to the Dec 5 model, the abundance of O is reduced and that of
S increased, which is the behaviour expected as deeper and deeper regions of
the ejecta are probed. The reduced velocity is the reason for the de-blending
of many features. There are two major shortcomings in the model: 1) the
position of the peak at 5500\AA\ is not correctly reproduced. This is clearly
the consequence of the absorption at 5400\AA\ not being correctly reproduced,
so that the `optical depth window' starts too far to the blue. This feature is
not well reproduced in the spectra of SNe~Ia either, so it is quite likely that
some Fe-group lines are missing or their strength is not correct. 2) The \SiII\
line is too narrow, or rather the very triangular shape of the blue side of the
absorption cannot be reproduced. Only another \SiII\ line is active in that
region (\SiII\ 5979\AA, observed near 5800\AA), but it is too weak. In any case
it would be difficult to explain the observed profile with just one line, so we
really have no explanation for what might affect the spectrum.

We have also modelled the spectrum using our modified density distribution
(Figure 4, continuous line). As we could expect, since the near-photospheric
region is not affected by the modification, at this epoch the two synthetic
spectra are essentially identical. The spectrum computed with the modified
outer density profile has $B = 17.73$ and $V = 16.66$. The bolometric
correction is still large, $BC = 0.20$, and $M_{Bol} = -16.87$. It is very
interesting that the model luminosity, as obtained from fitting the spectrum,
is actually larger on Dec 17 than on Dec 5. The bolometric light curve appears
then quite different from both the $V$ and the $B$ curves. The large bolometric
correction is now caused mostly by the red part of the spectrum, with the
synthetic $R$ band still rising so that the bolometric light curve rises slowly
to a rather delayed maximum. Near-IR photometry would have been useful to
confirm this prediction.

\section{The 24 December 1997 spectrum}

The 24 Dec spectrum is the first of a set of three spectra modelled in this 
paper which extend the time coverage beyond the near-maximum epochs considered
also in Paper I. The 24 Dec spectrum is shown in Figure 5. It does not extend
far to the red, but in the blue it has a good S/N down to 3700\AA, thus
covering the \CaII\ H\&K region. The epoch of the spectrum is 35 days after
explosion, for which we chose Nov 20. The spectrum has $v = 16.8$, a drop of
0.2 mag in only one week from Dec 17, and lies therefore well in the declining
part of the $V$ light curve. The \SiII\ line is now quite narrow, and its core
has a velocity of 5400\kms. This and all other absorptions continue to move to
the red. Following the trend already noticeable on Dec 17, many small features
which earlier blended into the broad troughs are beginning to be resolved,
marking the progressive inward motion of the photospere. The broad absorption
near 6000\AA\ is the region that changed the most with respect to Dec 17. In
particular, a separate absorption is now visible at 5700\AA.

We modelled the spectrum using model CO100 with the outer modification
introduced as discussed above. As usual, $L$ and $v_{\rm ph}$ were selected so
as to get as good a fit as possible to the observations. The input parameters
for the best model, which is shown as the continuous line in Figure 5, were $t
= 34$ days, $\log L = 42.13$ (erg s$^{-1}$) and $v_{\rm ph} = 4900$\kms. Note
that $L$ has the same value as on Nov 29, indicating that the bolometric light
curve is now declining. The photospheric velocity is now very low, as is also
indicated by the \SiII\ line velocity, and is actually slightly below the inner
edge of the density distribution of model CO100 (cf. Paper I, Figure 3).
Therefore, in order to produce a reasonable spectrum we had to introduce a
second change to the CO100 density structure, extending it down to $v_{\rm
ph}$. We used $\rho \propto r^{-1}$ for this inner extension. This is only a
first approximation, but it is obvious that if there is a photosphere there
must be opacity right down to it, and this is not compatible with the `density
hole' predicted by model CO100 below 5300\kms. Since the photosphere is so
deep, the inclusion of the flat outer density profile derived from the earlier
spectra has no effect on the synthetic spectrum. Even when the flat outer
density is included, \OI\ 7774\AA\ and the \CaII\ IR triplet are predicted to
be separated in wavelength. This region is not covered on Dec 24, but our
prediction will be confirmed in the Jan 1 spectrum (see next section).  The
small inner extension to the density does not affect the total mass much, so
that at this epoch the total mass above the photosphere is 8.4\Msun.

A major change in this model is in the abundances. As shown in Figure 7 of
Paper I, the Oxygen-dominated envelope extends down to 6500\kms. On Dec 24 the
photosphere is well below this point, and Si, S and the Fe-group have larger
abundances. We fitted the abundances to obtain a best fit, and found that S is
the element whose abundance must increase most, at the expense of O. The
behaviour of the abundances as determined from spectrum synthesis is reviewed
in the discussion. Most line identifications are the same as on earlier epochs.
The absorption at 5700\AA\ is due to \NaI D. Overall, the synthetic spectrum is
a good match to the observed one. The major shortcoming is that the emission
peak near 6500\AA\ is not reproduced. Net emission may be present there,
although the wavelength does not match that of the expected line of [\OI]
6300\AA. There is a wavelength coincidence with H$\alpha$, but this is not
expected to be a strong feature in a SN~Ic ejecta.

The synthetic spectrum has $B = 17.83$ and $V = 16.71$. The bolometric
correction is still large, $BC = 0.32$, so that $M_{Bol} = -16.70$. Therefore,
it appears that bolometric maximum was reached near Dec 24, much later than $V$
or $B$ maximum. A lot of the flux is still released in the near-IR, in
particular in the \CaII\ IR triplet peak.

\section{The 1 January 1998 spectrum}

The next spectrum in our series is rather close in epoch to the previous one
but, with an epoch of 42 days, it captures a moment when the light curve begins
to settle on the tail. Although the S/N is only moderately high, the ample
wavelength coverage (3400 to 8900\AA) makes it an interesting spectrum to
model. The spectrum is shown in Figure 6. Comparing it to the Dec 24 spectrum,
several signs of development can easily be noticed: 
\begin{enumerate}
 \item  The ratio of the three peaks in the blue, at 4000, 4500 and 5500\AA\ 
is much steeper, indicating a shift of the flux towards the red which clearly 
must be the consequence of a reduced temperature. 
 \item  The absorption complex near 6000\AA\ has now split into two components. 
One is centred near 5700\AA\ and is due to \NaI D, which is now sufficiently 
strong and isolated because of the reduction in strength of neighbouring lines 
coming from more highly excited levels that it gives rise to its own P-Cygni 
emission, near 5900\AA. To the red of \NaI D the \SiII\ line is still strong, 
near 6250\AA, and is accompanied by a weaker absorption near 6100\AA, which was 
also present on Dec 24. 
 \item  Two absorptions are developing at about 6800 and 7100\AA. 
 \item  The \OI\ 7774\AA\ line and the \CaII\ IR triplet are now well
separated,  as was already predicted by our synthetic spectrum on Dec 24.  The
\OI\ line is  rather narrow, while the \CaII\ absorption is broader. This is a
consequence of  the different run of excitation with radius (excitation falls
more steeply for  O than for Ca, and so does the line optical depth).
 \item  Emission in the \CaII\ IR triplet is now very strong, much stronger
than the corresponding absorption: this indicates that net emission is
beginning to be an important contribution to spectrum formation. 
 \item  The line blueshift is further reduced: the \SiII\ line now has a
measured $v = 3600$\kms. This is now significantly smaller than than the inner
boundary of the original CO100 density distribution. Since the essentially
photospheric nature of the spectrum is proved by the progressive evolution of
all features, a viable model must include a low velocity extension to the
density profile.
\end{enumerate}
Therefore, we modelled the spectrum using as a starting point the CO100 density
distribution modified at high velocity as described earlier. This modification
has no effect on the emerging spectrum at these later epochs: the maximum
observable matter velocity, measured at the blue edge of the \CaII\ IR triplet 
is only 23000\kms. We extended the density distribution inwards to $v =
3000$\kms, and tried various density laws below 5000\kms. Our first conclusion
is that the density must be increasing inwards: if we use a decreasing density
a shell forms near 5000\kms which affects the lines; if the density is constant
below 5000\kms the value near the photosphere is too low and the lines are too
weak. On the other hand, if the density increases too steeply the synthetic
lines tend to be too sharp. As a best approximate solution we adopted a power
law with $\rho \propto r^{-1}$.

Our final model has parameters $t = 42$ days, $\log L = 42.0$ (erg s$^{-1}$), 
$v_{\rm ph} = 3600$\kms, which is close to the observed \SiII\ velocity. It is
interesting that the simultaneous reduction in both $L$ and $v_{\rm ph}$ means
that the electron temperature $T_e$ ranges between 4500 and 6500K, which is
close to the values it had on previous epochs. This explains why the naure of
most line features is unchanged throughout the evolution of the SN. The
synthetic spectrum is shown as the thin line in Figure 6. Overall, the quality
of the fit is rather good, confirming the correctness of our assumption. Major
shortcomings are: 1) the strength of the peak at 4500\AA. This is probably due
to having somewhat too much opacity in the UV. 2) The strength and position of
the peak near 5500\AA. This was a problem also on Dec 24 and we explained it as
the consequence of missing line strength near 5400\AA. 3) Flux is missing in
the  \CaII\ IR triplet emission. This is because net emission - following
collisional excitation - is not included in our code. The difference between
the observed and synthetic peak is to be ascribed to net emission, which is
responsible for about half of the total line flux.

Other parts of the spectrum are very well reproduced, \eg the \OI\ and \CaII\ 
IR triplet absorption, the structure of the \SiII\ line region, where the weak
absorption at 6000\AA\ is attributed to \FeII\ 6148\AA\ and the emission near
6500\AA\ is now correctly reproduced, confirming that the peak on 24 Dec was
probably not [\OI] 6300\AA. The weak absorptions at 6800 and 7100\AA\
correspond to lines of \OI\ 7001\AA\ and to several \FeII\ lines, respectively,
but the synthetic lines are too weak. \FeII\ lines are becoming stronger as the
abundance of Fe-group elements is higher, while that of O is further decreased.

The inward extension of the density affects the ejecta mass, adding about
1.5\Msun\ between 3000 and 5000\kms, but it only has a small effect on the
kinetic energy, adding only $2 \cdot 10^{50}$erg in that velocity shell. The
synthetic spectrum has $V = 17.16$, which compares well with the observed $V =
17.3$. Synthetic $B = 18.41$, showing that the spectrum is rather red. The $R$
magnitude drops less than both $B$ and $V$, and the bolometric correction is
still large, $BC = 0.22$, so that $M_{Bol} = -16.35$, but this neglects the net
emission in the \CaII\ IR triplet, so the actual value is probably smaller.

\section{The 26 January 1998 spectrum}

This is the last spectrum available, corresponding to an epoch well on the
light curve tail, with $V \sim 18.0$. The spectrum, shown in Figure 7, shows a
clear evolution from Jan 1. In particular, the flux peak has moved to the red
quite noticeably, indicating a temperature drop. Looking at the spectrum in
detail, one can notice that the features in the blue have changed little, and
so has the region furthest to the red, although the \CaII\ IR triplet emission
is now stronger. On the other hand, the region around 6000\AA\ has changed
quite significantly. The flux peak has moved from 5500 to 6000\AA, the
absorptions at 5700\AA\ (\NaI D) and at 6200\AA\ are much stronger. The \SiII\
line has essentially disappeared, all that is left being a weak feature near
6400\AA, which is too red to be identified with the \SiII\ doublet. In fact, a
\SiII\ velocity cannot be measured at this epoch. Still, many features have
persisted from earlier epochs, and these have shifted further to the red,
indicating the presence of significant amounts of material at very low velocity
in the ejecta.

The overall nature of the spectrum is still photospheric, but the incidence of
net emission is higher than on Jan 1, which will limit our ability to fit the
spectrum using the MC code. We used the density distribution discussed in the
previous section, including both the outer $\rho \propto r^{-4}$ part above
25000\kms, which has no effect on the synthetic spectrum, and the inner $\rho
\propto r^{-1}$ extension between 3000 and 5000\kms, but we had to extend the
density distribution further inwards, so that we could obtain an appropriate
temperature and spectrum. In keeping with the hydrodynamical prediction of a
`density hole' in the centre of the ejecta, we tried to use the largest
possible value of the density power law index allowed by the quality of the
synthetic spectrum. We selected a model based on the overall fit, but we could
not reproduce the observations as well as on previous epochs. We could
accomodate an innermost density law $\rho  \propto r$ below 3000\kms, which is
an attempt to include an inner `hole' in density. This extension adds a further
0.5\Msun\ to the ejecta mass, but only less than $10^{50}$erg to the explosion
kinetic energy. Our best fit spectrum, shown in Figure 7, has parameters $t =
67$ days, $\log L = 41.80$ (erg s$^{-1}$) and $v_{\rm ph} = 1950$\kms. The
temperature is significantly lower than on Jan 1, with $T_{\rm e}$ ranging from
6000 to 4000K. Because of the rather flat density structure at low velocity,
the temperature is almost constant ($T_{\rm e} \sim 6000$K) at $v \leq
6000$\kms, which is where the spectrum is formed. This helps to keep the lines
sufficiently broad. The abundances are similar to those of Jan 1, but Oxygen is
further reduced and S is increased, reflecting the expected trend as smaller
and smaller velocities are sampled.

Admittedly, the synthetic spectrum is not a very good fit, but almost all the
observed features are at least reproduced. The two emission peaks at 4600 and
5500\AA\ are much too strong in the model, which affects the synthetic $V$
magnitude. The observed value is $V = 18.0$, and the synthetic spectrum has $V
= 17.85$. The $B$ flux is also somewhat overestimated at $B = 19.06$. Possible
reasons for the excessive strength of the synthetic peaks at 4600 and 5400\AA\
were already given in the previous section. In the $V$ region, an absorption
near 5600\AA\ is also not reproduced. A similar feature was also observed on
Jan 1, near 5650\AA, but it was weaker then, while it was absent on Dec 24. 

The absorption might be attributed to \HeI\ 5876\AA\ at a velocity of about
13000\kms. He may be expected to be found in the ejecta as a leftover of the
star's He-shell. If He is distributed in a shell, He atoms may only be excited
at a rather advanced epoch, when the $\gamma$-rays and positrons from the decay
of \Cofs\ can penetrate the ejecta and reach the shell. However, the velocity
implied by the  position of the absorption is low for a hypotetical He shell,
and the apparent shift of the feature between Jan 1 and Jan 26 does not support
this scenario either.  Alternatively, He can be produced with 56Ni in the
deepest layer as the result of alpha-rich freezeout (see Paper I, Figures 5 and
6).  In this case He would be found together with Co/Fe, so the He velocity
could take any value depending on the 56Ni mixing. However, in this case
non-thermal excitation of He would take place immediately, and we should see
the \HeI\ line as soon as the photosphere reaches the layer where He is
present. If He is located at about 14000\kms, we should have seen the \HeI\
line as early as Dec 5.  The above problems notwithstanding, as long as no
other candidate identification is available He must be regarded as at least a
possibility. Observatons of \HeI\ 10830\AA\ would certainly help settle this
issue. In any case, given the moderate strength of the line, the He mass
involved may not have to be large.

The model predicts a moderately strong absorption at 6300\AA, attributing it to
a blend of the \SiII\ doublet and, mostly, of \FeII\ 6417, 6456\AA. This
feature is present also in the observed spectrum, but it is much weaker. The
reason for the discrepancy is clearly the emission at 6300\AA, which must
signal the onset of the corresponding [\OI] line. Another net emission, which
is not reproduced by the model, is visible near 7300\AA. This is clearly \CaII]
7292, 7324\AA. Together with the \CaII\ IR triplet, these are the only three
net emissions in the optical spectrum. This is a typical situation for SNe~Ic
at these epochs (cf. SN~1987M at 60 days after maximum, Swartz \etal 1993).

Comparing the spectra of SN~1997ef and SN~1987M, apart from the obviously
broader lines in SN~1997ef, it is possible to note that in SN~1987M the \FeII\
lines are weaker and that the candidate \HeI\ line is not seen. SN~1987M is
thought to have produced roughly as much \Nifs\ as SN~1997ef (0.15\Msun;
Nomoto, Filippenko, \& Shigeyama 1990, Paper I), but this could be an
overestimation. Classical, low energy SNe~Ic are not expected to eject any He,
but this might not be the case for SN~1997ef, and it certainly is not for the
other hypernova, SN~1998bw, where \HeI\ 10830\AA\ is observed (Patat \etal
2000). On the other hand, at a comparable epoch the [\OI] 6300\AA\ emission is
much stronger in SN~1998bw than in SN~1997ef, supporting the idea that
SN~1998bw came from a more massive progenitor. Further detailed study of the He
(and O) content of hypernovae would certainly be worthwhile.

The synthetic spectrum has a small $BC$ (0.01), and $M_{Bol}= -15.87$.
The actual value of $BC$ may be even smaller, and possibly negative, since the
model $V$ flux is certainly overestimated by about 0.15 mag, and the net
emission in the \CaII\ IR triplet is not included in our estimate. Therefore,
the value of $M_{Bol}$ has an uncertainty of at least 0.2 mag.

\section{The explosion properties of SN~1997ef} 

As we have discussed above at great length, modelling the time evolution of the
spectrum of SN~1997ef has revealed two main inconsistencies with the explosion
model CO100, which we used in Paper I to reproduce both the light curve and the
near-maximum spectra. The basic parameters of the observed and synthetic 
spectra are recapped in Table~1. 

Firstly, the high line velocities observed near maximum in several strong
lines, notably the \CaII\ IR triplet and the \SiII\ doublet, could not be
reproduced with the rather steep density profile of CO100, but require a
flatter outer density law, $\rho \propto r^{-4}$, at $v > 25000$\kms.

Secondly, the fact that the January spectra are still predominantly
photospheric in nature, showing low line velocities ($v < 5000$\kms) is not
compatible with the presence in CO100 of an inner `density hole'. This `hole'
was the result of depositing the kinetic energy at a radius in the progenitor
which results in the ejection of exactly the amount of \Nifs\ necessary to 
reproduce the SN tail luminosity. Matter below that radius, known as the `mass
cut', is assumed to fall back onto the compact remnant. Clearly, material must
exist at low velocities, and we found that an inner extension below 5000\kms
with a $\rho \propto r^{-1}$ density law down to 3000\kms\ and a $\rho \propto
r$ law below that, to reproduce the inner `density hole', is the best overall
solution, although we still cannot get perfect fits of those very late epochs,
even ignoring the net emission which is clearly present in some lines.

In Figure 8 we show the density profile of the original CO100 model and our
modified one. The outer density extension adds significanlty to the kinetic
energy ($\sim 10^{52}$erg), but only marginally to the ejecta mass (0.25\Msun),
while the inner extension adds about 1.65\Msun\ to $M_{ej}$ but has a
negligible effect ($\sim 2 \cdot 10^{50}$erg) on the explosion kinetic energy.
Our modified model has a total $M_{ej} = 9.5$\Msun\ and $KE = 1.9 \cdot
10^{52}$erg, compared to $M_{ej} = 7.6$\Msun\ and $KE = 8 \cdot 10^{51}$erg for
CO100. These new values reinforce the classification of SN~1997ef as a
hypernova, which is based on the $KE$ being larger than $10^{52}$erg.

Since we have a $\rho \propto r^{-4}$ density law on the outside, and an
essentially flat law deep inside, and since our density distribution was
derived starting from that of CO100, whose relic is the steep density gradient
in the intermediate part, one might argue that a $\rho \propto r^{-4}$ density
law might work well throughout the ejecta. We have tested such a possibility,
but although it works reasonably well for the outer part, at $v \sim
20000$\kms, it gives rise to very sharp-lined spectra at the later epochs, when
the photosphere is in the flat part of the ejecta, at $v < 10000$\kms. Also,
joining directly the inner flat part and the outer $\rho \propto r^{-4}$ part
beyond 10000\kms\ would give a SN with extremely large $M_{ej}$ and $KE$, and
the synthetic spectra for the epochs when the photosphere is near 10000\kms\ 
would have very deep lines. 

We have to ask ourselves the question what could give rise to the observed
deviation from the hydrodynamical model. The flat outer part may be due to mass
loss at high rate during the presupernova stages, or to a transition between
the Oxygen shell and an outer Helium shell, some evidence for which may be
present as a weak \HeI\ 5876\AA\ line in the two January spectra, although at a
velocity ($\sim 13000$\kms) which is smaller than what would be expected.
Branch (2000) offered a similar solution, and his ad hoc density
law was even flatter $\rho \propto r^{-2}$ than ours. As a result he offers an
estimate of the kinetic energy above 7000\kms\ as $3 \cdot 10^{52}$erg. Our
value above the same velocity is somewhat smaller but comparable, $1.8 \cdot
10^{52}$erg. Also, he suggests that the ejecta mass above 7000\kms\ is $\sim
6$\Msun, and our value is again smaller but comparable, $\sim 4$\Msun.

The location of the inner density cutoff, on the other hand, was determined
essentially by demanding that the correct mass of \Nifs\ be ejected from the
progenitor. Clearly, the estimate of the position of the cut-off was incorrect.
Although it is possible that this may be due to an incorrect calculation of the
progenitor's structure, it is more tempting to attribute the problem to some
asymmetry in the explosion. If the explosion was asymmetric, similar to what
has been suggested for the hypernova SN~1998bw in order to explain its
connection with a GRB, it is quite possible that most \Nifs\ was produced near
the beam axis, while away from that axis burning would be less efficient, and
would terminate at intermediate elements such as Si or S (Maeda et al. 2000).
Clearly one then has to place the mass cut deeper in order to achieve the
require ejected mass of \Nifs. This is also supported by our derived abundance
distribution, which suggests that the S abundance is still increasing at the
lowest velocities sampled, about 2000\kms. The abundance distribution as
derived from our models is displayed in Figure 9, but we must keep in mind
that, since we used homogenised compositions above the photosphere, all sharp
composition boundaries are smoothed out. On the other hand, no clear spectral
evidence - such as sudden changes in the properties of the lines - is visible
for strict abundance stratification.

The presence of an inner density core was also suggested in Paper I to explain
the observed deviation of the synthetic light curve computed with model CO100
from the observations at advanced phases: the synthetic light curve has a
steeper slope than the observed one, indicating that $\gamma$-ray trapping is
more efficient than in the model. This could be achieved if \Nifs\ was
distributed deeper than it is in CO100.

\section{A new light curve model} 

In order to verify our findings, we used our derived density and abundance
distribution and computed a synthetic light curve in spherical symmetry.  We
used two different codes, one based on Monte Carlo methods and the other, a
radiation hydrodynamics code, which was used in the light-curve computation in
Paper I.

The Monte Carlo code is based on the simple code developed for SNe~Ia 
described in Cappellaro \etal (1997). $\gamma$-ray deposition in the expanding
ejecta is computed following the random walk of gamma-ray photons adopting a
constant $\gamma$-ray opacity $\kappa_{\gamma} = 0.027$cm$^2$g$^{-1}$. Once
they deposit their energy, the $\gamma$-rays are assumed to generate optical
photons on the spot. The random walk of the optical photons through the ejecta
is then also followed in Monte Carlo, assuming a constant optical opacity
$\kappa_{opt}$. The Monte Carlo scheme is able to model efficiently the random
walk of photons through the ejecta - and thus to take into account the delay
between the emission of a photon and its escaping from the SN nebula, which
determines the initial shape of the light curve.  The value of $\kappa_{opt}$
depends on the temperature and composition, and it is possible that it changes
with time. Thus we had to find a convenient approximate value for this
particular model of SN~1997ef. Since $\kappa_{opt}$ affects the light curve 
essentially near maximum, we found that a value $\kappa_{opt} = 0.08$
cm$^2$g$^{-1}$ gives a good fit to the light curve around maximum. This value
is smaller than for SNe~Ia, as might be expected since line opacity is much
stronger in the Fe-dominated ejecta of a SN~Ia.

The synthetic light curve in Monte Carlo is compared to the one obtained with
the radiation hydrodynamics code. As described in Paper I, the code solves a
multi-frequency radiative transfer equation in the fluid's comoving frame,
using the Feautrier method iteratively with an approximate Lambda operator. 
The energy equation of gas plus radiation and the radiation momentum equation
are coupled to the transfer equation in order to follow the evolution of gas
temperatures. The code also uses average opacities, and the values used were 
the same as in the Monte Carlo code.

The synthetic bolometric light curves are shown in Figure 10. Except for a
slight difference around the maximum, which is probably to be ascribed to 
Monte Carlo noise, the codes produce very similar light curves. Therefore,
it is safe to use the simple Monte Carlo code in calculating bolometric
luminosities if a reasonable opacity is chosen.

The synthetic light curves reproduce our derived bolometric curve reasonably
well around maximum. The $V$ light curve can consequently be reproduced if the
bolometric correction as resulting from our synthetic spectra is taken into
account. In both codes a \Nifs\ mass of 0.13\Msun\ was used. In order to
achieve a rapid rise of the light curve, consistent with our dating, we had to
distribute \Nifs\ homogeneously throughout the ejecta. Both light curves
reproduce the maximum quite well, but their decline on the tail is much too
steep. The observed $V$ light curve after about day 60 appears to follow the
\Cofs\ decline rate, implying constant and complete deposition of about
0.07\Msun\ of \Nifs.  However, our codes predict for those advanced epochs a
deposition fraction of less than 0.5 and steadily decreasing with time.

In Figure 11 we compare the observed velocity of the \SiII\ doublet with the
run of $v_{ph}$ as derived from the two light curve codes and with the values
used as input for the calculation of the synthetic spectra. The two codes are
in good agreement with one another, although the MC values are consistently
larger than those of the radiation hydro model. This is probably due to
different zoning in the two codes. However, the observed points are lower than
both models, and the difference increases with the epoch. The velocities used
for the spectral calculations, on the other hand, start lower than the observed
values but slowly approach them.  This is understandable, as it can be expected
that at early times the \SiII\ line forms above the photosphere since it is
very strong. Later, as the line becomes weaker, it forms closer and closer to
the photosphere. The fact that the value of $v_{ph}$ tends to flatten out at
advanced epochs, and does not reproduce the observations, is not new (see, \eg,
Iwamoto \etal 1998 on SN~1998bw), and it probably indicates that the assumption
of a gray photosphere is not good at advanced epochs. However, the difference
at early phases cannot be due to that effect. One possibility is that we have
somewhat overestimated the density at the highest velocities in our spectral
calculations, where we were guided more by the \CaII\ IR triplet than by the
\SiII\ doublet. Note that the velocity derived from model CO100 in Paper I was
much smaller than the observed values. Another possibility is that our
assumption of a gray opacity again fails in those outermost layers, as shown
perhaps by the fact that only the \CaII\ IR triplet is active there. Note that
all of these remarks apply to a spherically symmetric situation.

Another inconsistency between the light curve and spectral calculations is that
if we integrate the Ni mass as the sum of the abundances of Fe, Co and Ni in
the various synthetic spectra we obtain an Fe-group mass of only $\sim
0.04$\Msun above 2000\kms, which is much less that what is necessary to power
the light curve.  It is unlikely that our spectrum synthesis may have given
such a large error, especially since the Fe lines are already very strong in
most spectra. 

So we have a siuation where at early times we need more surface \Nifs\ to 
power the light curve than we see in the spectra, while at late times the
\Nifs\ we have used to reproduce the light curve at peak does not deposit 
sufficiently to reproduce the light curve slope and brightness. The late light
curve appears to require the complete deposition of $\sim 0.07$\Msun\ of \Nifs.
The two problems may be independent, but they may also be related. 

One possibility to explain the late light curve behaviour is that, in a
spherically symmetric scenario, additional \Nifs\ ($\sim 0.05$\Msun) could be 
located deep in the ejecta, below 2000\kms. If the density is sufficiently high
there the $\gamma$-rays produced by the decay of this \Nifs\ may deposit
completely. The presence of very low velocity ejecta would further reduce the
mass of the expected compact object, and may be in contradiction with the
distribution of \Nifs\ in models of the progenitor evolution and explosion.
This would not affect the situation at early times. 
 
Another possibility is that the ejecta may not be spherically symmetric, in
either mass or abundance distribution, or both. If some \Nifs\ is ejected at
low velocities together with other elements in some direction, the
$\gamma$-rays produced by its decay chain could be efficiently trapped even at
advanced epochs. The optical photons produced by the thermalisazion of
$\gamma$-rays and positrons in this hypothetical high density region would only
be able to escape at advanced phases, and so they would have little effect on
the early time spectra but they could power the late light curve.  At the same
time, some material has been ejected at high velocity, possibly in a jet-like
form, to reproduce the sharp rise of the light curve by high-velocity \Nifs\
and to explain the broad lines of \SiII\ and \CaII\ ($\sim 20,000$\kms) at
early epochs. In this scenario, the uniform mixing of \Nifs\ which was used to
reproduce the sharp rise of the light curve might refer only to the \Nifs\
ejected in the jet, while the low velocity \Nifs\ would enter play only later.

This scenario might help explaining the discrepancy between the light curve and
spectroscopic \Nifs\ masses. If most of the \Nifs\ is ejected in a preferred
direction (\eg in the jet that might be observed as a GRB if the viewing angle
is favourable), this \Nifs\ may power the early light curve by depositing its 
$\gamma$-rays in the neighbouring ejecta, which have lower velocity, higher
density and a smaller Fe-group abundance. However, if we view the event from an
angle sufficiently far from the jet direction, the spectrum we see would be
produced in a region where the abundances are different from those found in the
jet -- hence the smaller spectroscopic mass of \Nifs.  This is a possibility
for SN~1997ef, since there is no positive identificatin of a GRB counterpart to
the SN event.

The possibility that the explosion was asymmetric is in line with existing
models of energetic supernova explosions linked to GHRB's (MacFadyen \& Woosley
1999, H\"{o}flich \etal 1999, Khokhlov \etal 1999, Maeda \etal 2000), and it
deserved further study with multi-dimensional radiation hydrodynamics codes.

\section{Discussion} 

We have shown in this paper how spectrum synthesis can be used not only to
verify explosion models of SNe, but also to improve on them. Given our
findings, it would now be very interesting to see hydrodynamic calculations of
the explosion, in one or more dimensions, which reproduce closely the observed
bolometric light curve. The derived density distribution, ejecta mass and
kinetic energy may or may not be close to what we have obtained
spectroscopically, but in any case such a model would have to be tested also
for its ability to reproduce the SN spectral evolution. Our study also shows
that fitting an observed $V$ curve with a synthetic bolometric one may be
somewhat misleading, since the bolometric correction can be significant.

We have discussed in \S 9 the reason for the difference between the density
distribution derived from spectral modelling and that obtained from 1D 
hydrodynamic calculations of the explosion of the stripped core of a massive
star, and remarked that the presence of significant amounts of ejected matter
at low velocity may be due to some asymmetry in the explosion. A similar
conclusion was reached for the other Type Ic hypernova, SN~1998bw, but in that
case it was based on somewhat different evidence, most importantly the
connection with a GRB (\eg Wheeler 2000).

The early-time spectra of SN~1998bw have even broader and more blended lines
than those of SN~1997ef, but a clear evolution towards lower line velocities
and less line blending is not seen as clearly as in SN~1997ef, while the
relatively early development of nebular emission is seen in both objects. This
may be understandable since although the ejecta mass of SN~1998bw was larger
than that of SN~1997ef, even if we use our upwards revised value for the
latter, the kinetic energy of SN~1998bw appeared to be much larger (Nomoto
\etal 2000). Branch (2000) in a spectroscopic study similar to the one he
performed for SN~1997ef, suggested an extremely large $KE$ for SN~1998bw ($6
\cdot 10^{52}$erg above 7000\kms, \ie a factor 2 larger than his value for
SN~1997ef above the same velocity). We have performed a preliminary analysis of
SN~1998bw using the technique described in this paper, and find a similarly
large value (Nakamura \etal 1999).

If SN~1997ef was also a highly asymmetric explosion, although weaker in energy
and ejecting a smaller mass than SN~1998bw, why was a GRB not positively
detected? The most likely possibilities are that either the jet-axis of 
SN~1997ef was not oriented exacly along the line-of-sight, or that a weaker
explosion energy reduced the beaming so that a GRB was not formed. Both
scenarios might be able to explain many of the observed features of SN~1997ef,
in particular the lower expansion velocities and the smaller measured $KE$.
When we apply a spherically symmetric model, we estimate $KE$ by integrating a
1D model around a sphere. Therefore, if the SN is observed on or very close to
the jet-axis, like SN~1998bw, $KE$ could be grossly overestimated H\"oflich
\etal 1999). On the other hand, the difference between the velocity of the
spectral lines in the two SNe is real, and so is the difference in luminosity
at late times.  That SN~1998bw produced nore \Nifs\ is also confirmed by the
estimate for the \Nifs\ mass ($\sim 0.6$\Msun) obtained from the nebular lines,
which does not depend much on the asymmetry (see Danziger \etal 1999; Nomoto et
al. 2000).  Therefore, even though inclination may be a factor, that there is
some intrinsic difference between the two objects appears unavoidable, at least
as long as the relative distance estimate is reliable.

Finally, we note that another SN~1997ef-like Type Ic hypernova was recently
observed, SN~1998ey, whose spectra appear to be identical to those of
SN~1997ef, and for which there is also no observed GRB counterpart (Garnavich
\etal 1998). This coincidence is striking, and calls for the accumulation of
more data on Type Ic hypernova candidates. 

{\bf Acknowledgements} 
This work has been supported in part by the grant-in-Aid for
Scientific Research (12640233, 12740122) and COE research (07CE2002) 
of the Ministry of Education, Science, Culture and Sports in Japan.
It is a pleasure to thank P.Garnavich and J.Danziger for useful communications
and the referee, D.Branch, for constructive remarks.


\begin{deluxetable}{rrccccccccccccc}
\scriptsize
\tablenum{1}
\tablecaption{Parameters of the synthetic spectra}
\tablehead{\colhead{date} &
\colhead{epoch} &
\colhead{$L$} &
\colhead{$v_{ph}$} &
\colhead{$v({\rm \SiII})$} &
\colhead{$\log \rho_{ph}$} &
\colhead{Mass} &
\colhead{$T_{eff}$} &
\colhead{$T_{bb}$} &
\colhead{$B_{mod}$} &
\colhead{$V_{mod}$} &
\colhead{$V_{obs}$} &
\colhead{$BC$} &
\colhead{$M_{mod}$} \nl
\colhead{ } &
\colhead{days} &
\colhead{erg s$^{-1}$} &
\colhead{km s$^{-1}$} &
\colhead{$R_{\odot}$} &
\colhead{g cm$^{-3}$} &
\colhead{\Msun} &
\colhead{K} &
\colhead{K} & & & & &  \nl}
\startdata
29 Nov &  9 & 42.13 & 15500 & 19072 & -12.65 & 1.05 & 6123 & 7725 &
 17.45 & 16.74 & 16.7 & 0.29 & -16.600 \nl
 5 Dec & 15 & 42.15 &  9500 & 13962 & -12.30 & 3.38 & 6128 & 9447 &
 17.35 & 16.58 & 16.5 & 0.40 & -16.650 \nl
17 Dec & 27 & 42.20 &  7500 &  8011 & -12.67 & 5.16 & 5291 & 6648 &
 17.72 & 16.65 & 16.6 & 0.20 & -16.775 \nl
24 Dec & 34 & 42.13 &  4900 &  5378 & -12.71 & 7.56 & 5602 & 7602 &
 17.83 & 16.71 & 16.8 & 0.32 & -16.500 \nl
 1 Jan & 42 & 42.00 &  3600 &  3775 & -12.78 & 8.48 & 5426 & 6830 &
 18.41 & 17.16 & 17.3 & 0.22 & -16.250 \nl
26 Jan & 67 & 41.80 &  1950 &   --  & -13.51 & 8.97 & 5232 & 5701 &
 19.06 & 17.85 & 18.0 & 0.01 & -15.775 \nl
\enddata
\end{deluxetable}


\newpage
\begin{figure*}
\plotone{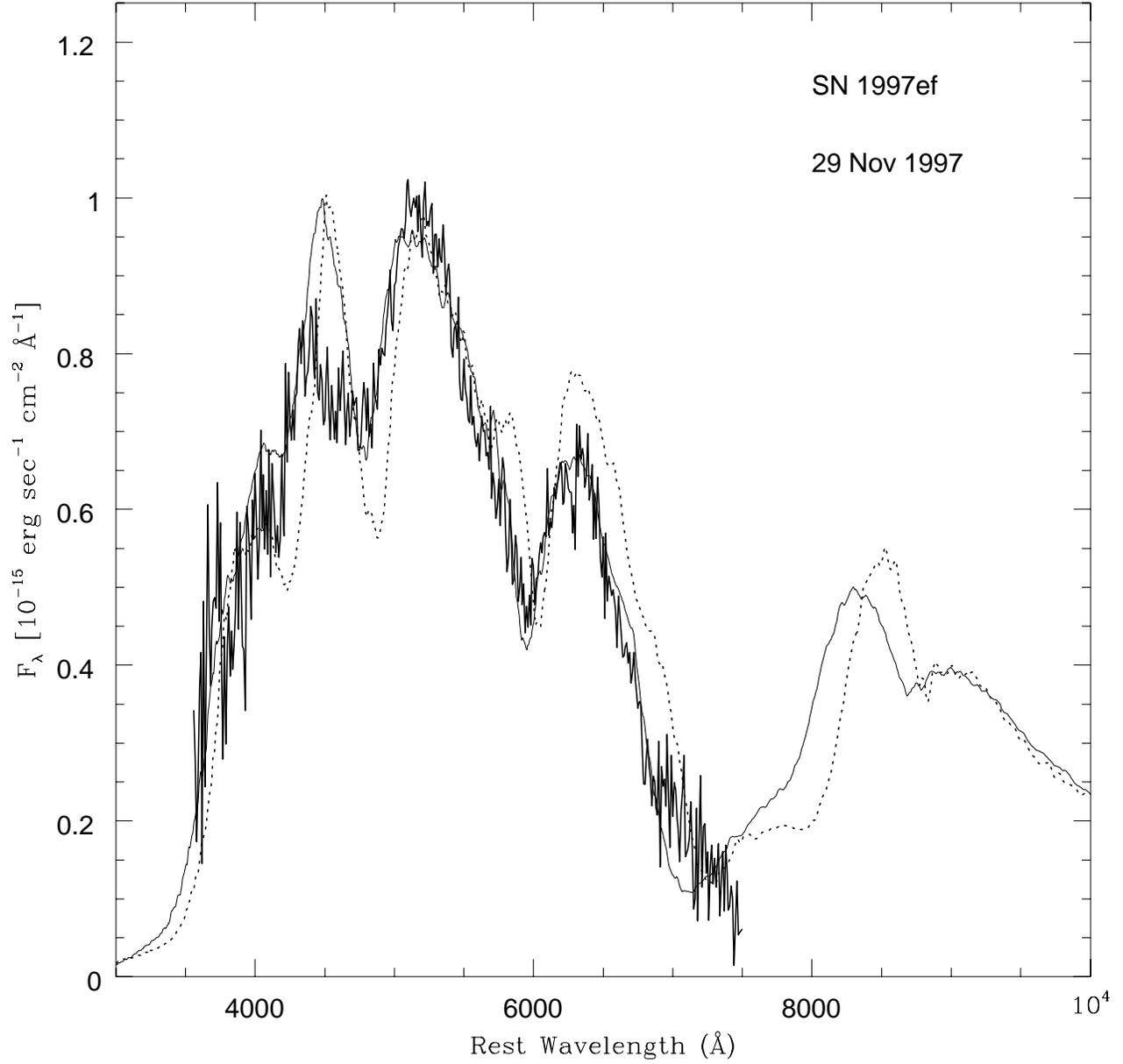}
\figcaption[.ps]{The observed, smoothed spectrum of SN 1997ef on Nov 29, 1997 
(thick line), compared to two synthetic spectra computed with the CO100 ad hoc 
density structure. The fully drawn thin line is a spectrum computed for $t=9$ 
days, while the dashed line is a spectrum computed for $t=11$ days. }
\end{figure*}

\newpage
\begin{figure*}
\plotone{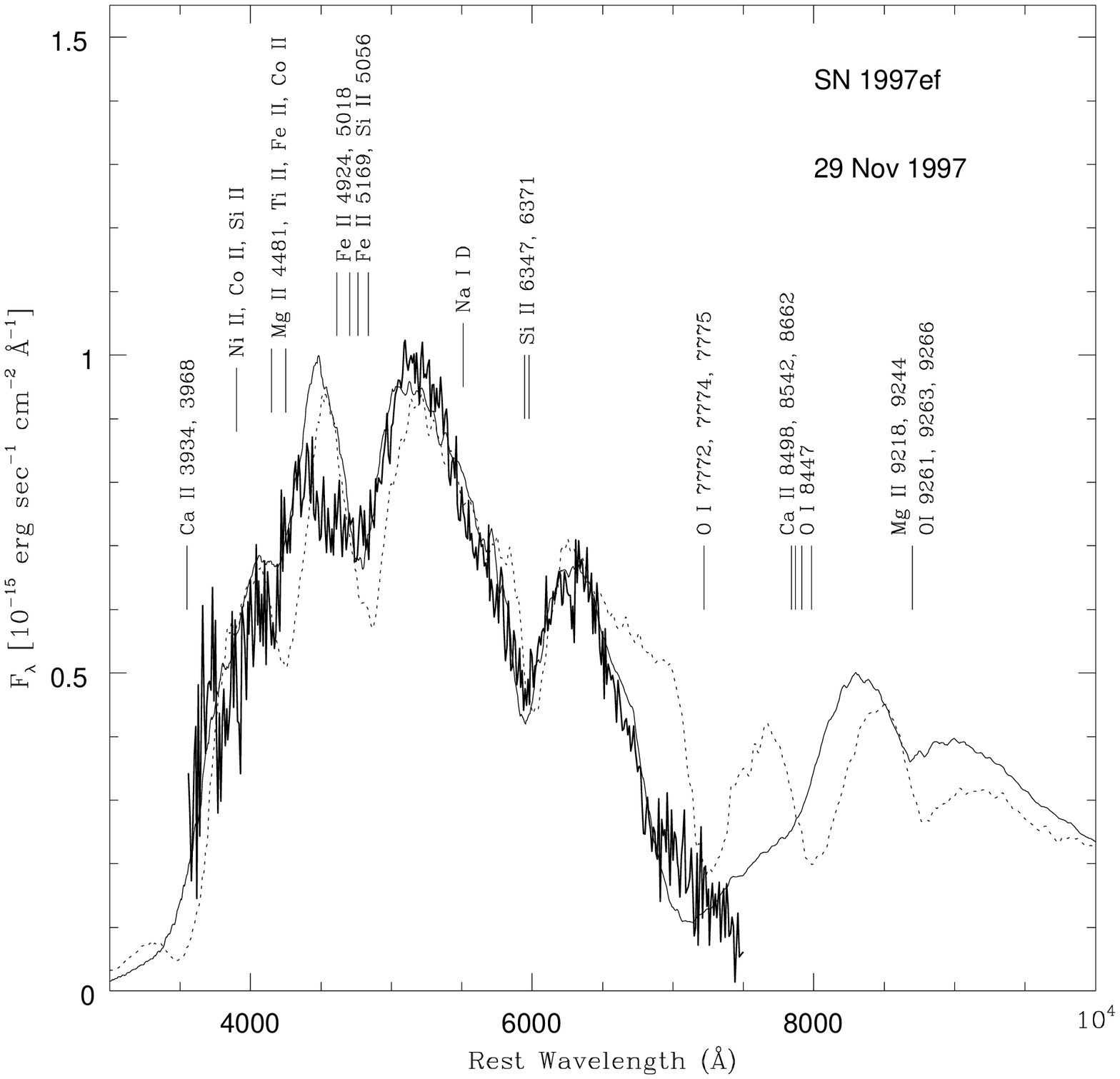}
\figcaption[.ps]{The observed, smoothed spectrum of SN 1997ef on Nov 29, 1997 
(thick line), compared to two synthetic spectra computed for $t=9$ days. 
The dashed line is a spectrum computed with the original model CO100, while the
fully drawn thin line is a spectrum computed with the modified outer density
described in the text. }
\end{figure*}

\newpage
\begin{figure*}
\plotone{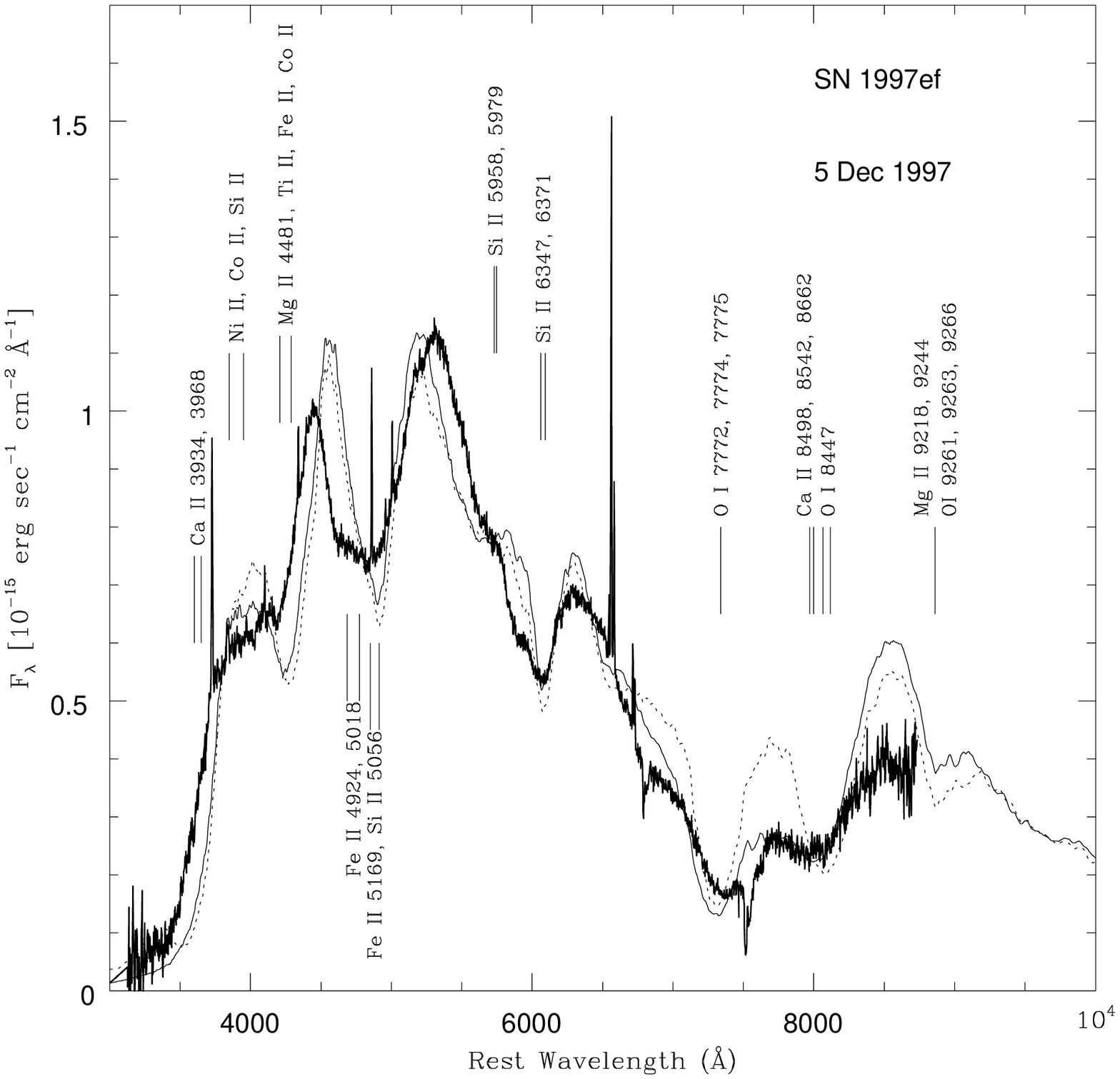}
\figcaption[.ps]{The observed spectrum of SN 1997ef on Dec 5, 1997 (thick
line), compared to two synthetic spectra computed for $t=15$ days. 
The dashed line is a spectrum computed with the original model CO100, while the
fully drawn thin line is a spectrum computed with the modified outer density
described in the text. }
\end{figure*}

\newpage
\begin{figure*}
\plotone{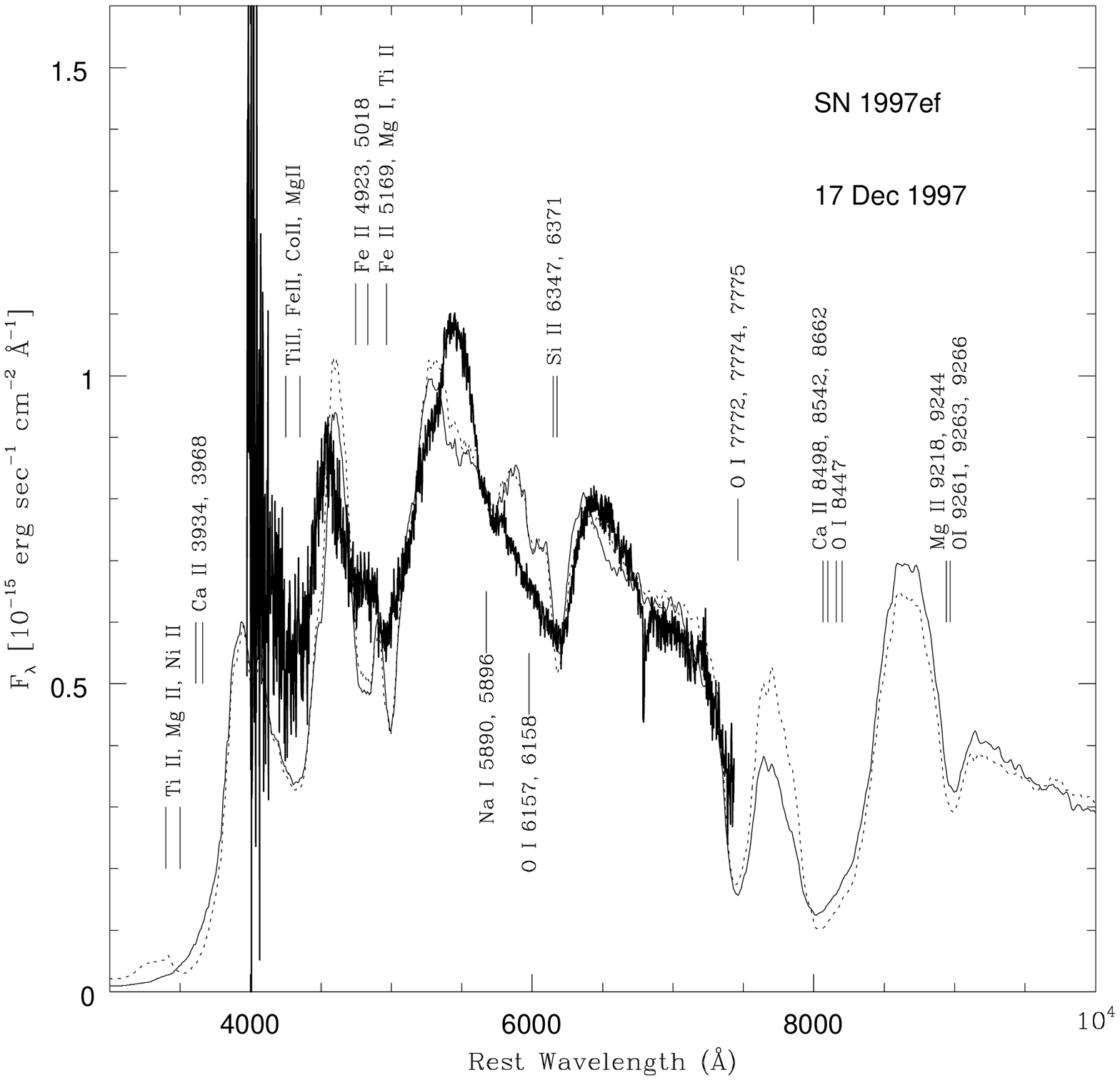}
\figcaption[.ps]{The observed spectrum of SN 1997ef on Dec 17, 1997 (thick
line), compared to two synthetic spectra computed for $t=27$ days. 
The dashed line is a spectrum computed with the original model CO100, while the
fully drawn thin line is a spectrum computed with the modified outer density
described in the text. }
\end{figure*}

\newpage
\begin{figure*}
\plotone{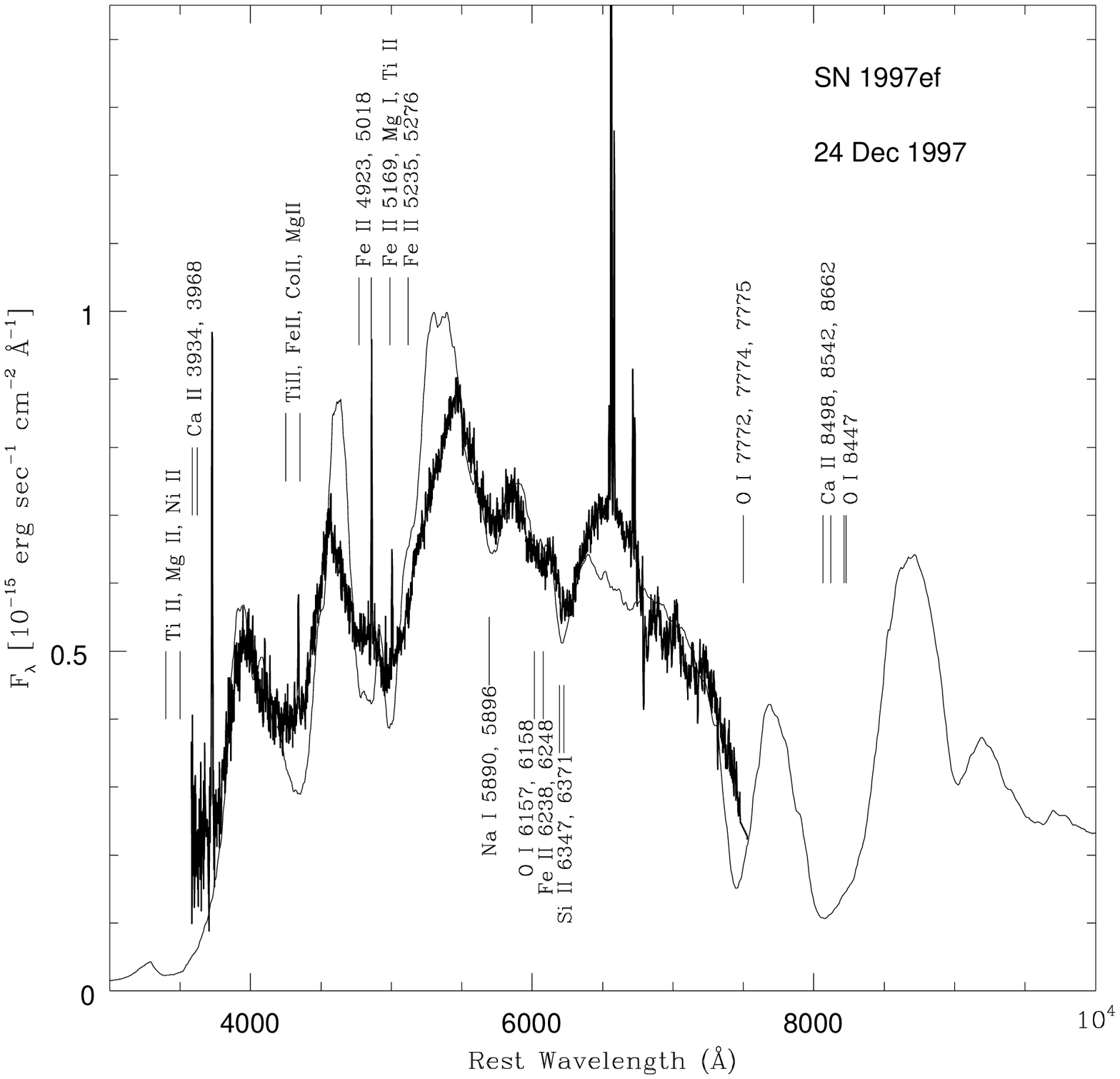}
\figcaption[.ps]{The observed spectrum of SN 1997ef on Dec 24, 1997 (thick
line), compared to a synthetic spectra computed for $t=34$ days using the 
modified density described in the text (thin line). }
\end{figure*}

\newpage
\begin{figure*}
\plotone{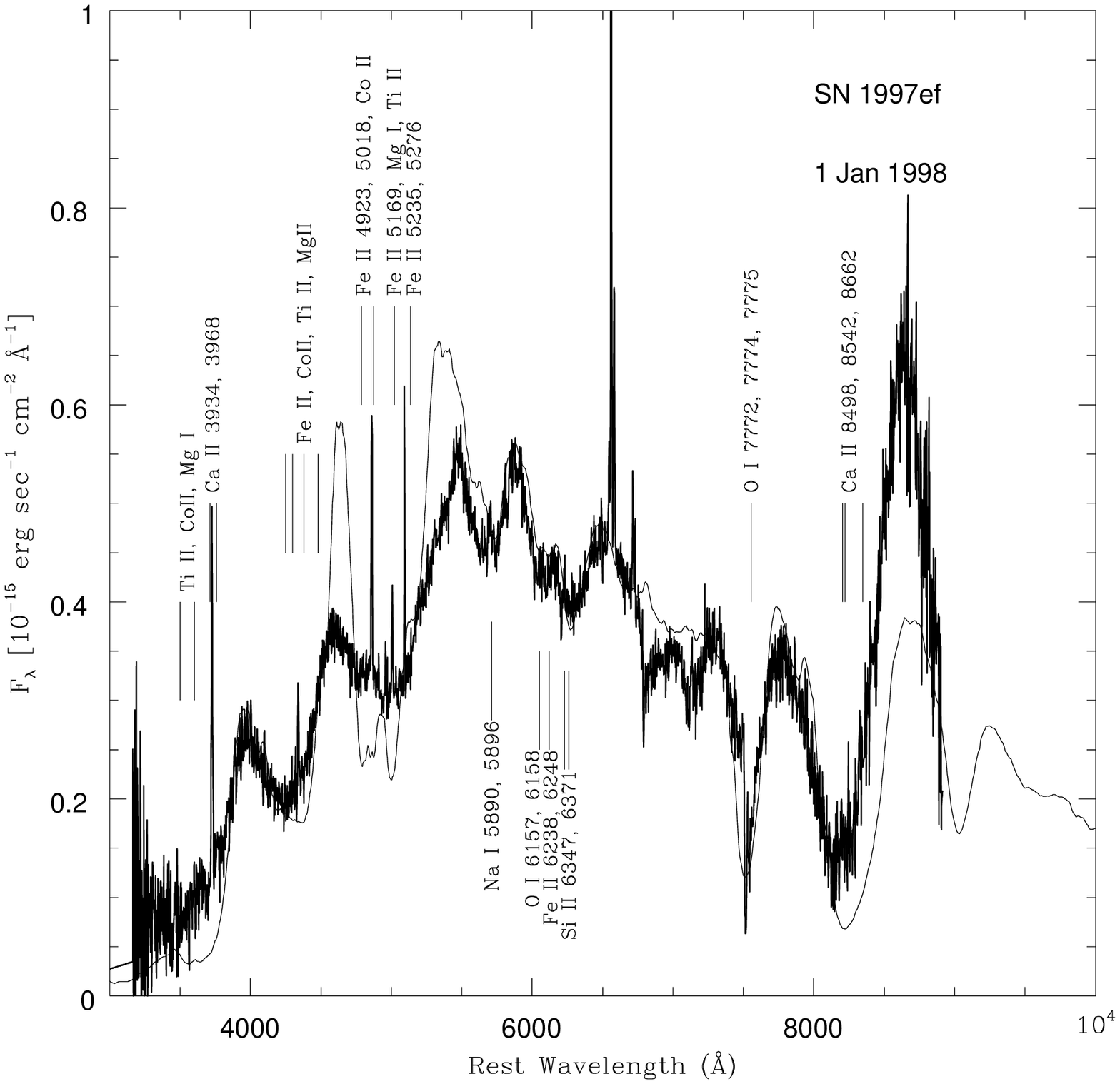}
\figcaption[.ps]{The observed spectrum of SN 1997ef on Jan 1, 1998 (thick
line), compared to a synthetic spectra computed for $t=42$ days using the 
modified density described in the text (thin line).  }
\end{figure*}

\newpage
\begin{figure*}
\plotone{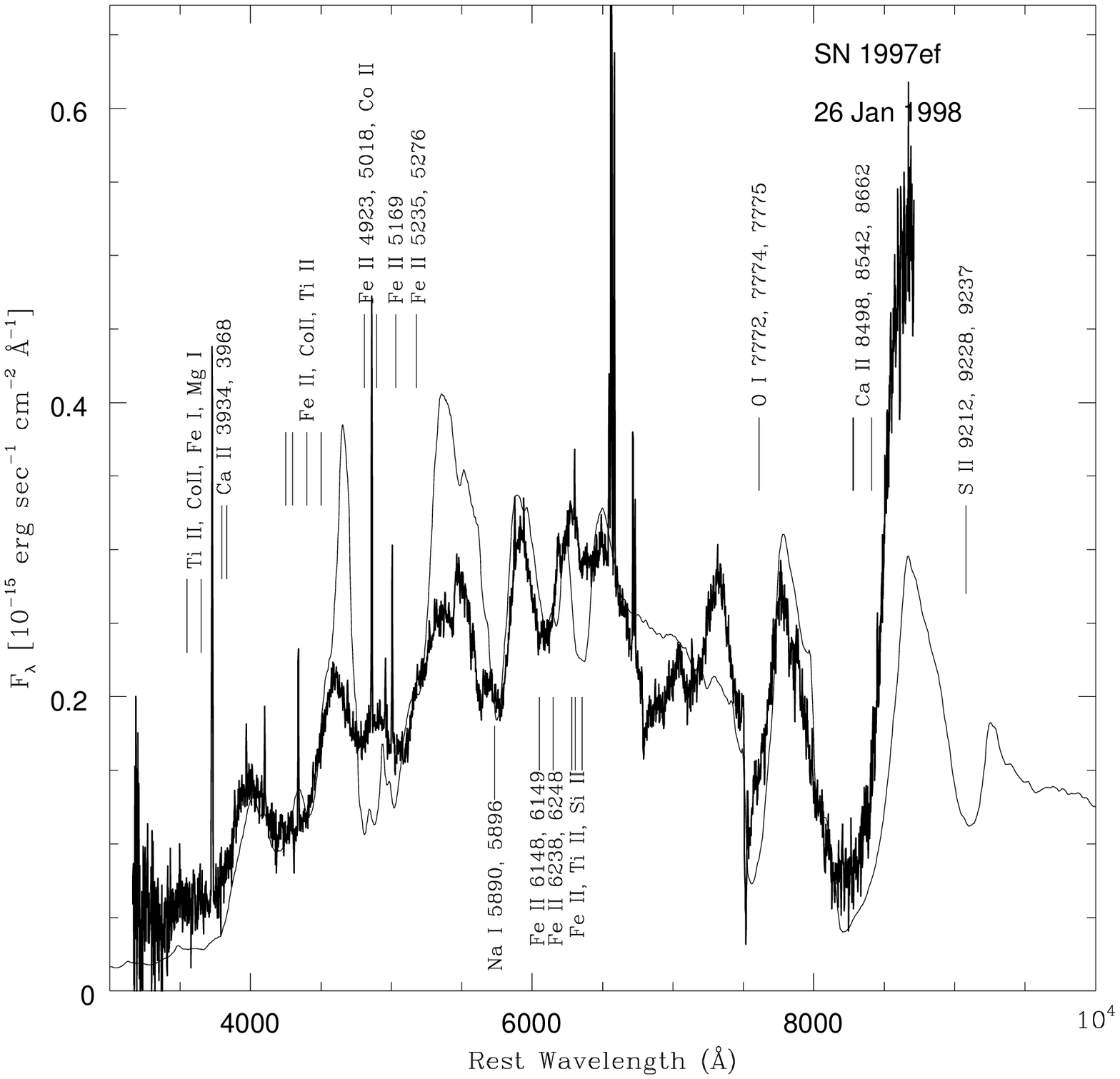}
\figcaption[.ps]{The observed spectrum of SN 1997ef on Jan 26, 1998 (thick
line), compared to a synthetic spectra computed for $t=67$ days using the 
modified density described in the text (thin line). }
\end{figure*}

\newpage
\begin{figure*}
\plotone{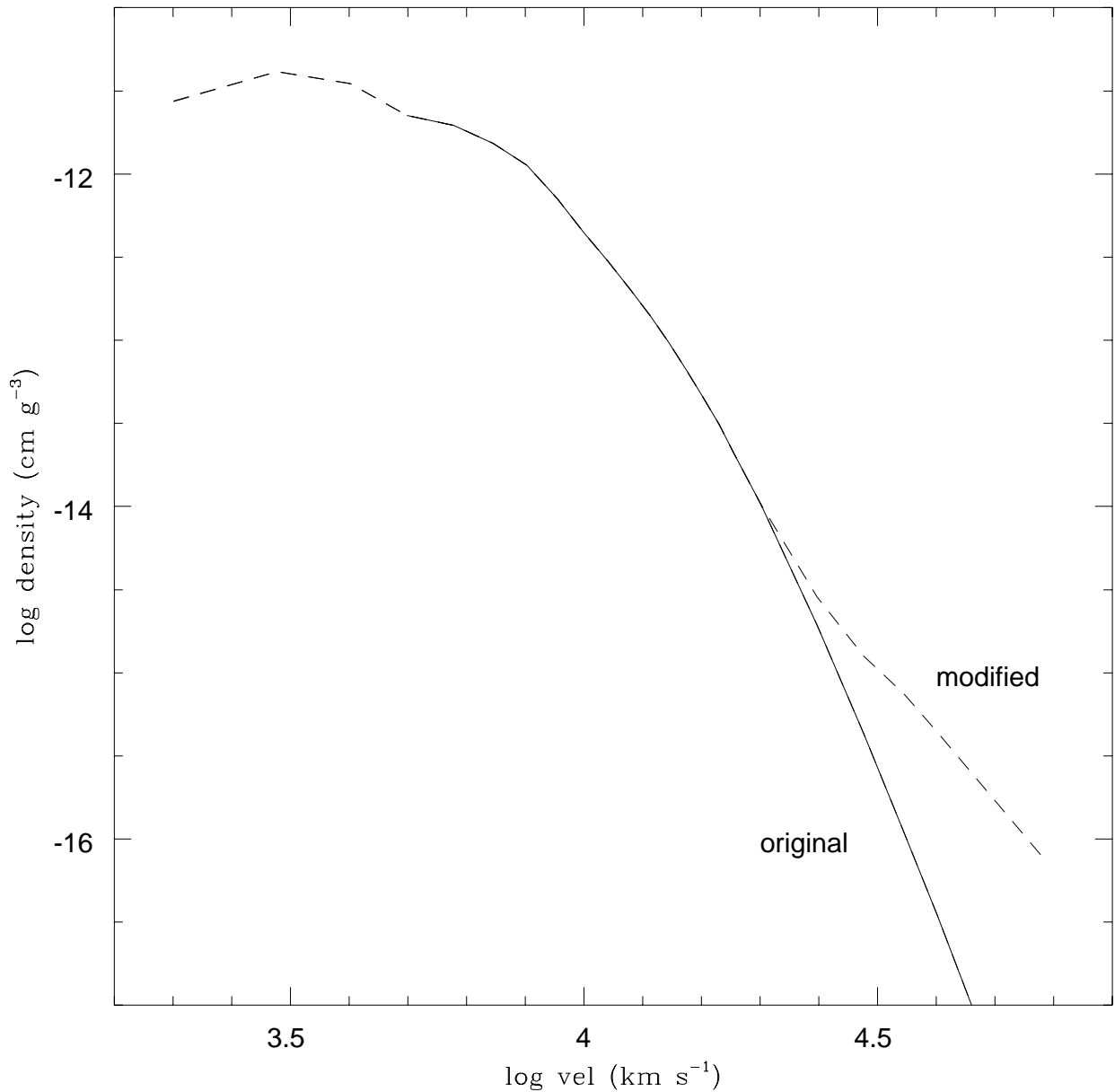}
\figcaption[.ps]{The original density structure of the hydrodynamical model 
CO100 (continuous line) and the modifications introduces to improved the
spectral fits (dashed line). The outer part of the modified density structure
has a power law index $n=-4$, while the inner extension has $n=1$ between
$v=3000$ and $v=5000$ \kms and $n=-1$ below $v=3000$ \kms. }
\end{figure*}

\newpage
\begin{figure*}
\plotone{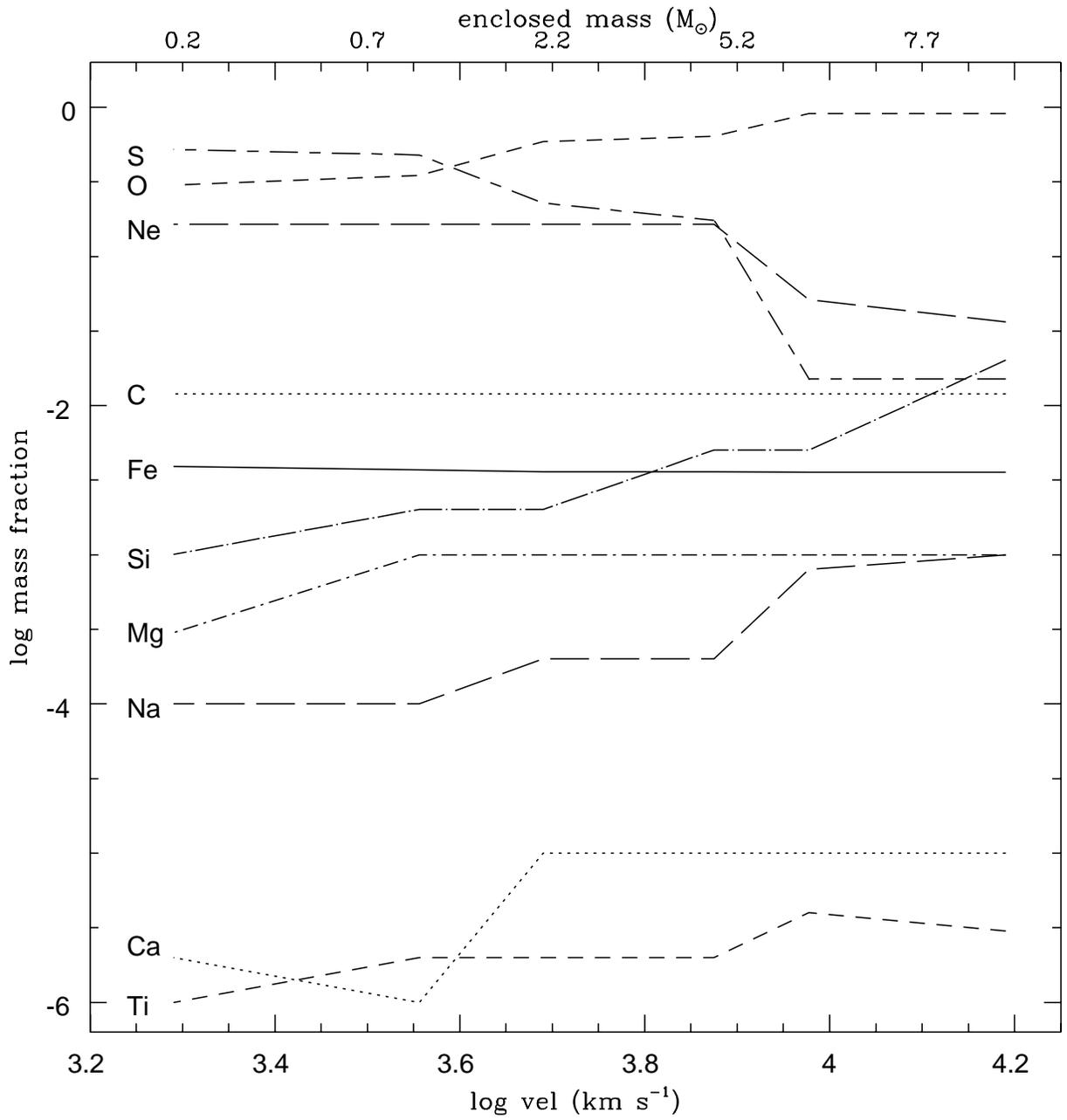}
\figcaption[.ps]{Composition structure in the ejecta of SN 1997ef as derived
from spectrum synthesis.  }
\end{figure*}

\newpage
\begin{figure*}
\plotone{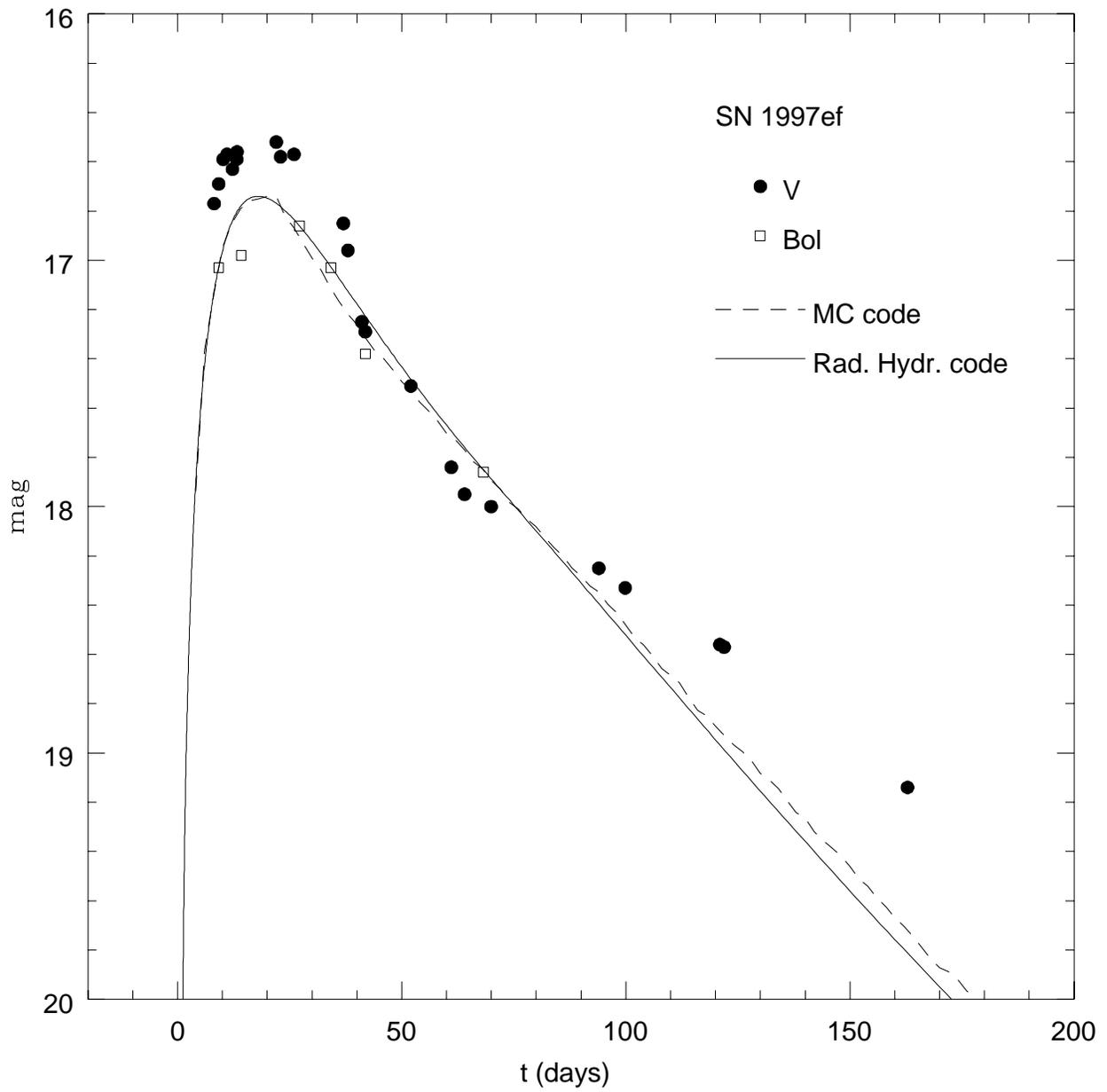}
\figcaption[.ps]{The synthetic bolometric light curves computed with the Monte
Carlo and the radiation hydrodynamics codes using the density distribution
derived from our spectral fits, compared to the quasi-bolometric light curve of
SN~1997ef. }
\end{figure*}

\newpage
\begin{figure*}
\plotone{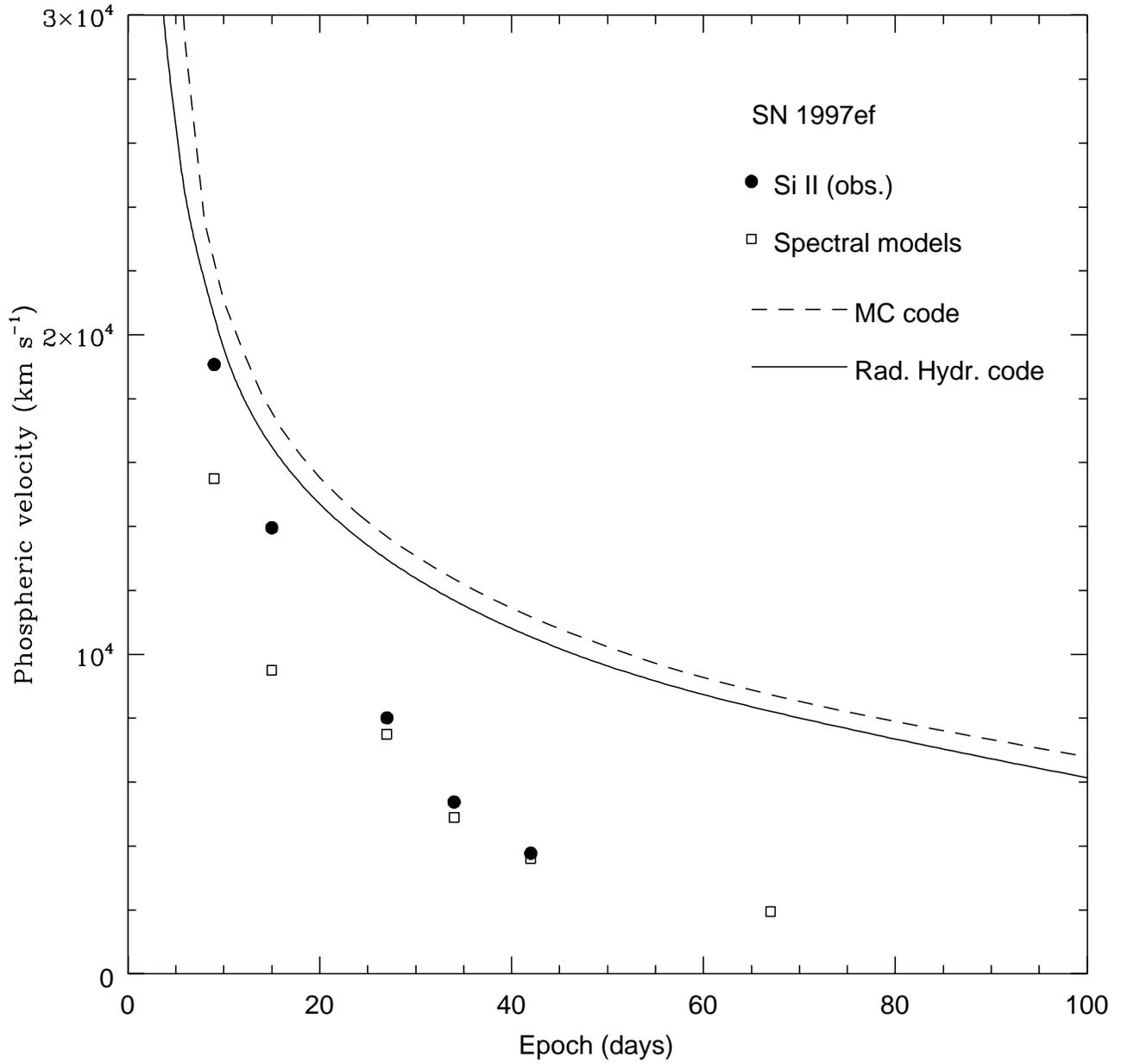}
\figcaption[.ps]{The evolution of the velocity of the observed \SiII\ doublet
and of the photospheric velocity as computed by the two light curve codes and
as derived from the spectral calculations. }
\end{figure*}


\end{document}